\documentclass[a4paper,11pt]{article}
\pdfoutput=1

\usepackage{jheppub} % for details on the use of the package, please see the JHEP-author-manual
\usepackage{siunitx}
\usepackage{booktabs} % To thicken table lines
\usepackage{amsmath}
\usepackage{braket}
\usepackage{multirow}
\usepackage[normalem]{ulem}
\usepackage{slashed}

\newcommand{\N}{{\cal N}}

\newcommand{\beq}{\begin{equation}}
\newcommand{\eeq}{\end{equation}}
\newcommand{\be}{\begin{equation}}
\newcommand{\ee}{\end{equation}}
\newcommand{\bea}{\begin{eqnarray}}
\newcommand{\eea}{\end{eqnarray}}

\newcommand{\el}{e^{-}}
\newcommand{\pos}{e^{+}}

\title{An updated view on the ATOMKI nuclear anomalies}

\author[]{Daniele Barducci and}
\author[]{Claudio Toni}
\affiliation[]{Universit\`a degli Studi di Roma la Sapienza, Piazzale Aldo Moro 5, 00185, Roma, Italy}
\affiliation[]{INFN Section of Roma 1, Piazzale Aldo Moro 5, 00185, Roma, Italy}
\emailAdd{daniele.barducci@roma1.infn.it}
\emailAdd{claudio.toni@uniroma1.it}

\abstract{In view of the latest experimental results recently released by the ATOMKI collaboration, we critically re-examine the possible theoretical interpretation of the observed anomalies in terms of a new BSM boson $X$ with mass $\sim17\;$MeV.
To this end we employ a multipole expansion method and
give an estimate for the range of values of the nucleon couplings to the new light state in order to match the experimental observations.
Our conclusions identify the 
axial vector state as the most promising candidate, while other spin/parity assignments seems disfavored for a combined explanation.
This results is however based on an order of magnitude estimate for the, currently unknown, axial nuclear matrix element of the $^{12}$C transition, that needs then to be evaluated before being able to draw a definite conclusion.
Intriguingly, an axial vector state can also simultaneously accommodate other experimental anomalies, {\emph{i.e.}} the KTeV anomaly in $\pi^0 \to e^+ e^-$ decay while 
being compatible with the conflicting measurements of the anomalous magnetic moment of the electron $(g-2)_e$ and other constraints on the electron couplings of the $X$ boson.
The PADME experiment will completely cover the relevant region of the parameter space, thus allowing for a strong test of the existence of the $X$ particle.

}

\begin{document} 
\maketitle

\section{Introduction}\label{intro}

The possibility that New Physics (NP) will manifest itself in the form of light and weakly coupled new states is nowadays raising more and more interest. This is mainly due to the null results from the LHC in searching for signs of TeV scale beyond the Standard Model (BSM) physics motivated by, {\emph{e.g.}}, the standard paradigms of compositeness or supersymmetry. While high$-p_T$ searches will continue to investigate these scenarios and, in the case of no discovery, further constrain them, it is of paramount importance to exploit any other experiment that can test the validity of the SM and ultimately falsify it. In this respect low energy and/or high intensity experiments provide an important probe. 

Among the various processes that can be investigated, rare nuclear transition can provide a good handle to observe NP appearing at the MeV scale, since they can significantly be affected by BSM physics even if this is very weakly coupled.
A nuclear transition occurs when an excited nucleus decays into a lower energy level of the same nucleus. Within the SM only electromagnetic (EM) interactions can mediated nuclear transition, which can mainly proceed through the following channels
\begin{itemize}
\item $\gamma-$emission, where the nucleus decays emitting a real photon,
\item Internal Pair Creation (IPC), where the nucleus emits a virtual photon which then decays to an $e^+e^-$ pair.
\end{itemize}

In recent years the ATOMKI collaborations has reported various anomalous measurements in the IPC decays of excited $^8$Be~\cite{Krasznahorkay:2015iga,Krasznahorkay:2018snd}, $^4$He~\cite{Krasznahorkay:2019lyl,Krasznahorkay:2021joi} and, more recently, $^{12}$C~\cite{Krasznahorkay:2022pxs} nuclei. These anomalies appear as bumps for both the invariant mass and the angular opening of the $e^+e^-$ pairs and have 
a high statistical significance, well above $5\sigma$. The ATOMKI collaboration has proposed to interpret them as due to the on-shell emission of a new boson $X$ from the excited nuclei, subsequently decaying to an $e^+e^-$ pair. The best fit mass for the hypothetical new particles is estimated to be $\sim 17\;$MeV.
Although to this day no independent confirmation of these results has arrived, given the multitude of processes in which these anomalies have been observed the ATOMKI results have attracted a considerable attention from the particle physics community. Many theoretical interpretation of the $X$ boson in terms of a new scalar or vector degree of freedom have been put forward, possibly unaccounted SM effects have been investigated and experimental searches have been proposed and/or are taking data with the goal of further investigating this anomaly.

In view of the latest experimental results recently released by the ATOMKI collaboration, we critically re-examine the possible theoretical interpretation of the anomaly in terms of a new BSM state. 
To this end we employ a multipole expansion method and
give an estimate for the range of values of the nucleon couplings to the new light state in order to match the experimental observations.
 We will focus on the $^8$Be and $^4$He anomalies and comment on how the measurement of the anomalous signal in $^{12}$C transitions impact our results. Our conclusions identify the 
axial vector state as the most promising candidate, while other spin/parity assignments seems disfavored for a combined explanation.
However  the axial nuclear matrix element of the $^{12}$C transition is currently unknown and, as we will show, our findings regarding the compatibility of an axial vector candidate with the $^{12}$C  anomalous transition
are based upon an order of magnitude estimate. Before being able to draw a definite and solid conclusion, the relevant matrix element must be evaluated.
Intriguingly, an axial vector state can also simultaneously accommodate other experimental anomalies, {\emph{i.e.}} the KTeV anomaly in $\pi^0 \to e^+ e^-$ decay 
while 
being compatible with the conflicting measurements of the  anomalous magnetic moment of the electron $(g-2)_e$ and other constraints on the electron couplings of the $X$ boson. The PADME experiment will completely cover the relevant region of the parameter space, thus allowing for a strong test of the existence of the $X$ particle.

The paper is structured as the following.
In Sec.~\ref{sec:review} we review the anomalous measurement of the ATOMKI experiment and the theoretical interpretation proposed so far. In Sec.~\ref{sec_sig} we describe the multipole expansion formalism while in Sec.~\ref{sec_X17} we present the results of the decay rates of the $^8$Be, $^4$He and $^{12}$C resonances for the various spin-parity assignment of the $X$ boson. We present our results in Sec.~\ref{sec:res} and we then conclude in Sec.~\ref{sec:conc}. We also add some appendices with more technical details. In App.~\ref{sec:app_num_table} we report some tables with the numerical values used throughout our analysis.
In App.~\ref{app:sec_EM} we apply the multipole formalism to the electromagnetic case of real $\gamma$ emission and IPC, while in App.~\ref{qton} we derive the effective nuclear couplings from the $X$ boson couplings to quarks. In App.~\ref{appBev} we present how our results change by considering also the $^8$Be(17.64) excited state. In App.~\ref{NWAApp} we report useful formul\ae~for the cross section of nuclear resonance production. Finally in App.~\ref{sec_pheno} we collect the experimental bounds for the spin-1 case relevant for our study.

\section{The ATOMKI anomaly}\label{sec:review}

The ATOMKI experiment~\cite{Gulyas:2015mia} consists in a proton beam colliding a target nucleus $A$ at rest, with the aim of producing an excited nucleus $N^{*}$ and measure its IPC transition to a ground state $N$, {\emph{i.e.}}
\be
p+A \rightarrow N^{*} \rightarrow N + e^+ e^- \ .
\ee
The list of nuclei used by ATOMKI and their main properties are reported in Tab.~\ref{tab:kin:1} and Tab.~\ref{tab:kin:2} in App.~\ref{sec:app_num_table}.
In 2015 the ATOMKI group studied the IPC decay channel from the $^{8}\text{Be}(18.15)$ and $^{8}\text{Be}(17.64)$ excited energy levels of Beryllium nuclei~\cite{Krasznahorkay:2015iga}. To populate the two states, a beam of protons was prepared in order to collider with target $^7$Li nuclei at rest.  By varying the energy of the incident proton beam, the collaboration was able to scan across the $^8$Be resonances. As a results they observed an anomalous peak corresponding to an opening angle for the $e^+ e^-$ pairs of $\sim \ang{140}$ for the IPC correlation distribution, in striking contrast with the QED prediction of a rapidly falling one. This has been interpreted as due to the decay of a short-lived neutral particles decaying into an $e^+ e^-$ pair, which would produce the observed peak at large angles. The observed deviation had a significance of $6.8\sigma$.
In order to confirm the anomalous origin of the signal, the collaboration repeated the measurement varying the energy of the incident proton beam. They found that the anomaly disappeared off the resonance peak, leading to the conclusion that it was probably originated by the decay of the $^8$Be excited energy level. The best fit mass for the hypothetical $X$ neutral boson has been estimated to be $m_{X} = 16.70 \pm 0.35 \pm 0.5\;$MeV, where the former uncertainty corresponds to the statistical error, while the latter to the systematic one. In 2018 the ATOMKI collaboration repeated the experiment with an improved setup~\cite{Krasznahorkay:2018snd}, which confirmed both the presence of the anomaly and the compatibility with the previous measurement.

\begin{table}[t!]
\begin{center}
\begin{tabular}{c|c|c|c|c|c}
$N$ & $N^{*}$ & $S^{\pi}$ & $I$ & $\Gamma(\text{keV})$ & $\Gamma_{\gamma}(\text{eV})$\\
\midrule
\midrule
$^{8}\text{Be}$ &  & $0^{+}$ & 0 & $5.57\pm0.25$ &  \\
\midrule
& $^{8}\text{Be}(18.15)$ & $1^{+}$ & $0^{*}$ & $138\pm6$ & $1.9\pm0.4$ \\
\midrule
& $^{8}\text{Be}(17.64)$ & $1^{+}$ & $1^{*}$ & $10.7\pm0.5$ & $15.0\pm1.8$ \\
\midrule\midrule
$^{4}\text{He}$ & & $0^{+}$ & $0$ & $\text{Stable}$ \\
\midrule
& $^{4}\text{He}(21.01)$ & $0^{-}$ & $0$ & $0.84$ & 0 \\
\midrule
& $^{4}\text{He}(20.21)$ & $0^{+}$ & $0$ & $0.50$ & 0 \\
\midrule\midrule
$^{12}\text{C}$ & & $0^{+}$ & 0 & Stable \\
\midrule
& $^{12}\text{C}(17.23)$ & $1^{-}$ & $1$ & $1150$ & $44$ \\
\end{tabular}
\end{center}
\caption{Spin-parity $J^{\pi}$ and isospin $I$ quantum numbers, total decay widths $\Gamma$ and $\gamma$-decay widths $\Gamma_{\gamma}=\Gamma(N^{*}\rightarrow\,N\,\gamma)$ for the nuclei used in the ATOMKI experiment: 
$^{8}\text{Be}$ \cite{Tilley:2004zz}, $^{4}$He \cite{Tilley:1992zz,Walcher:1970vkv} and $^{12}$C \cite{Kelley:2017qgh,Segel:1965zz} nuclei. Asterisks on isospin assignments indicate states with significant isospin mixing.}
\label{tab:nuclei}
\end{table}

Later in 2019 the ATOMKI group replicated the experiment using $^4$He nuclei~\cite{Krasznahorkay:2019lyl}, with the aim of searching for the anomalous signal in a difference source. The excitation energy was chosen to lie between two different resonances: the $^{4}$He$(20.21)$ state, with $J^{\pi}=0^{+}$, and the $^{4}$He$(21.01)$ state, with $J^{\pi}=0^{-}$. In this case the decay widths of the two excited states are large enough so that they can substantially overlap, so in the experiment both the excited states were populated, although off the resonance peak. In this case the IPC process was only possible for the $^4$He(20.21) states, while it's forbidden for the $^4$He(21.01), because of parity conservation in electromagnetic interactions. The group again found a rather sharp bump in the $e^+ e^-$ angular opening analogously to what has been observed in the $^8$Be case, with a significance equal to $7.2\sigma$. Interestingly, the peak was found to be located at an angle of $\sim \ang{115}$ which is compatible with the kinematics arising from the decay of the hypothetical $X$ boson with a best fit mass of $m_{X} = 16.98 \pm 0.16 \pm 0.2\;$MeV. Also the anomaly in the $^4$He channel was later confirmed by a second measurement, at different energies of the proton beam~\cite{Krasznahorkay:2021joi}.
More recently, the group has released a new analysis,
where the experiment has been replicated using now $^{12}$C nuclei~\cite{Krasznahorkay:2022pxs}. Also in this case an anomalous signal has been observed, with a peak at a larger value of the $e^+ e^-$ opening angle
$\sim \ang{150} - \ang{160}$, again compatible with the kinematic of the $X$ particle, with a best fit mass of $m_{X} = 17.03 \pm 0.11 \pm 0.20\;$MeV. In Tab.~\ref{tab:nuclei} 
we list all the ground and excited states considered in the ATOMKI analyses, together with their main properties: spin-parity assignment $S^\pi$, isospin $I$, total decay width  $\Gamma$ and $\gamma$-decay transition width $\Gamma_\gamma$.

Clearly, new and independent measurements are needed in order to confirm, or disproof, the results of the ATOMKI collaboration and test the consistency of the $X$ particle hypothesis. The MEG II experiment~\cite{MEGII:2018kmf} at PSI has the possibility to repeat the ATOMKI measurement on $^8$Be nucleus. At present time, the first dedicated data taking has recently been completed and data analysis is ongoing~\cite{Chiappini:2022egy}. A similar experiment is also being set up at the Montreal Tandem accelerator~\cite{Azuelos:2022nbu} with
data taking that should take place in early 2023, and at the Van-de-Graaff  laboratory~\cite{Cortez:2023ycv}.
Finally, also the PADME experiment~\cite{Raggi:2014zpa,Raggi:2015gza} in Frascati is in its data taking phase, and dedicated analyses will completely test the 
available parameter space for the coupling of the $X$ boson to electrons relevant for the explanation of the ATOMKI anomalies~\cite{Nardi:2018cxi,Darme:2022zfw}. 
 
These results and the interpretation given by the ATOMKI collaboration in terms of a new BSM particle comprehensibly attracted the attention of the theory community. However there also exist the possibility that the anomalous signal is due to unknown and/or underestimated SM effects. In this respect,  after the publication of the $^8$Be measurements, an attempt has been made in order to explain the anomaly with effects arising from known nuclear physics. In~\cite{Zhang:2017zap}  it has been proposed an improved nuclear physics model of the experiment, inspired by the so-called Halo Effective Field Theory (EFT) framework \cite{Hammer:2017tjm}, showing 
that the nuclear form factor needed to explain the anomaly suggests an unrealistic large length scale on the order of $10$ fm for the $^{8}$Be nucleus. Other Authors investigated the possibility of new exotic bound states~\cite{Chen:2020arr,Kubarovsky:2022zxm,Wong:2022kyg} as possible SM explanation of the ATOMKI anomaly while, 
on less exotic lines, it has been claimed in~\cite{Aleksejevs:2021zjw} that the experimental results can be reproduced within the SM by carefully considering the full set of next-to-leading-order corrections and the interference terms to the Born-level decay amplitudes, also proposing experimental improvements in order to test this hypothesis. All together it is fair to say that no firm explanation of the ATOMKI measurements in terms of SM effects has been established.

As  regarding possible BSM interpretation, the observation of the $^8$Be and $^4$He transitions restrict the $X$ boson to be either a vector, an axial vector or a pseudoscalar state, under the assumption of definite parity. These options have been all investigated in recent literature both from a model independent and/or effective parameterization and from a more ultraviolet (UV) completed perspective.
Among all the possibilities the one of a spin-1 boson stands out as an appealing one, since it could be related to a new symmetry of Nature. This scenario has been deeply  analyzed in~\cite{Feng:2016jff,Feng:2016ysn,Feng:2020mbt}, where, by working within an EFT framework, it has been found that a combined explanation of the $^8$Be and $^4$He anomalies in terms of a new vector states is possible, with the main constraint on this explanation coming from the search for a dark photon $\gamma_D$ in $\pi^0 \to \gamma\, \gamma_D$ decay by the NA48/2 collaboration~\cite{NA482:2015wmo}, whose non observation requires 
the $X$ boson to be {\emph{protophobic}}. 
Subsequently in~\cite{Zhang:2020ukq} it has been pointed out that the contribution from a {\emph{protophobic}} vector boson with mass around $17\;$MeV to direct proton capture processes, {\emph{i.e.}} to processes which do not proceed through an intermediate resonance, would be dominant with respect to the contribution from the resonant $^8$Be(18.15) state, in sharp contradiction with the experimental observation that the anomaly disappears off the nuclear resonance~\cite{Krasznahorkay:2015iga}. However a new experimental result from ATOMKI~\cite{Sas:2022pgm} claims to have observed the anomaly at different energies of the proton beams, opposite to their previous results. The group explains that this difference is due to a wrong estimate of the background in the previous analyses and then reopens the window for a {\emph{protophobic}} vector scenario.
The case of a pure axial vector has been investigated in~\cite{Kozaczuk:2016nma} where the authors applied a multipole expansion method to the anomalous nuclear decay rates and evaluated the related nuclear matrix elements by {\emph{ab-initio}} calculation using realistic nuclear forces. Their estimation concludes that the $X$ axial coupling to quark should be order ${\cal O}(10^{-4}-10^{-5})$ to explain the anomalous signal in the $^8$Be transition. 
Vector with mixed parity have also attracted attention, especially in the case of more specific BSM UV construction, see {\emph{e.g.}}~\cite{DelleRose:2018eic,Pulice:2019xel,Fayet:2016nyc}.
Lastly, the possibility of a light pseudoscalar has been considered in~\cite{Ellwanger:2016wfe}, where the authors made a rough estimation of the range of the values of the Yukawa couplings, assuming a nuclear shell model for the $^8$Be nucleus. The strongest constraints they reported come from flavor changing neutral current interactions as $K \to \pi X$, which however can be satisfied simultaneously explaining the ATOMKI $^8$Be results. The interesting possibility of the QCD axion being responsible for the ATOMKI anomalies has been entertained in~\cite{Alves:2017avw,Alves:2020xhf}, see also~\cite{Liu:2021wap}. Here the Authors focus on a axion candidate with dominant coupling to the first generation of SM fermions and piophobic, {\emph{e.g.}} with suppressed isovector coupling.
It has subsequently however been pointed out in~\cite{Hostert:2020xku} that for such a scenario a large pion decay rate for $\pi \to 3 X \to 3 e^+ 3 e^-$ of ${\cal O}(10^{-3})$ is expected, exceeding the SM double-Dalitz decay by a factor of thirty. There has been no direct measurement of this process so far, although it is reasonable to assume that such a large decay rate would have been noticed.

\subsection{Process kinematics}\label{sec_kin}

The ATOMKI anomalies show simple but well defined features, which are:
\begin{itemize}
\item the excesses are resonant bumps located at the same $e^+ e^-$ invariant mass for all the $^8$Be and $^4$He transitions,
\item the $e^+ e^-$ opening angles of the anomalous peaks are around $\ang{140}$, $\ang{115}$ and $\ang{155}-\ang{160}$, respectively, for the $^8$Be, $^4$He and 
$^{12}$C,
\item the anomalous signal in the $^8$Be transition have been observed only inside the kinematic region given by $|y|<0.5$, where $y$ is the energy asymmetry of the lepton pair, {\emph{i.e.}} the ratio between the difference and the sum of their energies.
\end{itemize}
As we review below, closely following earlier results appeared in~\cite{Feng:2016ysn,Feng:2020mbt},  these features are naturally explained by the hypothesis of resonant production of a new particle.

\begin{figure}[t!]
\begin{center}
\includegraphics[width=0.6\textwidth]{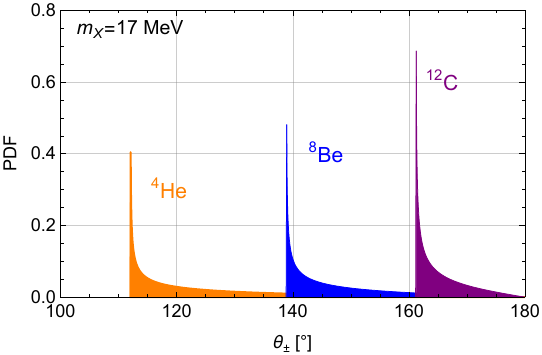}
\end{center}
\caption{Normalized distributions of the $e^+ e^-$ opening angles from the $^8$Be (blue), $^4$He (orange) and $^{12}$C (purple) nuclear transitions.}
\label{fig:pdf_th}
\end{figure}

In the experimental setup the target nucleus $A$ is at rest in the laboratory frame, while the proton beam energy $E_b$ is of the order of MeV, so that the colliding protons are mostly non relativistic.
The Center of Mass (CM) energy $E_{\rm CM}$ is then given by
\be
E_{\rm CM}=\sqrt{(m_{p}+m_{A})^{2}+2m_{A}E_{b}}\simeq m_{p}+m_{A}+\frac{m_{pA}}{m_{p}}E_{b} \ ,
\ee
where $m_{pA}= (m_{A}^{-1}+m_{p}^{-1})^{-1}$ is the reduce mass of the proton-target system. The experiment calibrates the beam energy in order to populate the $N^*$ state, which is produced almost at rest in the CM frame, and then measure its IPC transition to the ground state $N$.
Regardless on whether the resonance is (fully) populated, the collision between the proton and the target leads to the production of the $N$ nucleus via the emission of a boson of mass $m$, which could be a real or virtual photon or an hypothetical BSM particle. In the CM frame the boson energy $\omega$ is given by 
\be
\omega=\frac{E_{\rm CM}^{2}+m^{2}-m_{N}^{2}}{2E_{\rm CM}}\simeq E_{\rm th}+\frac{m_{pA}}{m_{p}}E_{b} \ ,
\ee
where the threshold energy $E_{\rm th} = m_p + m_A - m_N$ is the energy gap between the $N$ nucleus and the proton-target system. Note that the boson energy roughly only depends on the beam energy. Once produced the (real or virtual) boson decays into an $e^+ e^-$ pair, whose angular correlation is the main observable measured by the ATOMKI experiment. In the CM frame the total energy of the leptons is given by $\omega = E_+ + E_-$, where $E_+ (E_-)$ is the positron (electron) energy, while we label the opening angle between the leptons with $\theta_{\pm}$. The decay of the boson into the $\pos\el$ pair is controlled by the energy asymmetry
\be
y=\frac{E_{+}-E_{-}}{E_{+}+E_{-}},
\ee
while the opening angle is given by
\be
\label{thetapm}
\theta_{\pm}=\cos^{-1}\left(\frac{-1-y^{2}+\delta^{2}+2v^{2}}{\sqrt{(1-\delta^{2}+y^{2})^{2}-4y^{2}}}\right),
\ee
where $\delta=2m_{e}/\omega$ ($0<\delta<1$) and
\be
v=\sqrt{1-\left(\frac{m}{\omega}\right)^{2}}
\ee
is the boson velocity. 
\begin{figure}[t!]
\begin{center}
\includegraphics[scale=0.99]{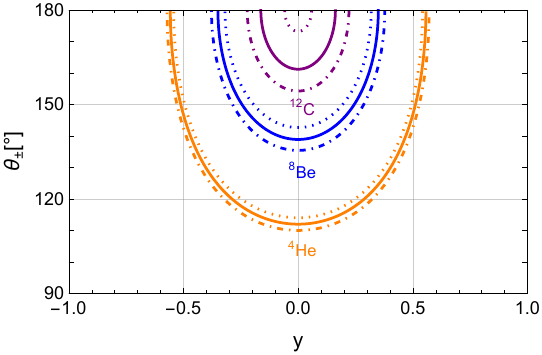}
\caption{Values of the $e^+ e^-$ opening angle $\theta_{\pm}$ as a function of the energy asymmetry $y$ for three values of the boson mass: $m_{X}=16.8$ MeV (dot-dashed line), $m_{X}=17$ MeV (solid line) and $m_{X}=17.2$ MeV (dotted line) for the cases of the $^8$Be, $^4$He and $^{12}$C transitions.}
\label{figt:kin:1}
\end{center}
\end{figure}
For an hypothetical $X$ boson with mass $m_{X}=17$ MeV, the maximum value of the energy asymmetry and the minimal value of the opening angle are respectively 
\be
y_{\rm max}\simeq v_{X} \hspace{2cm} \text{and} \hspace{2cm} \theta_{\pm}^{\rm min}\simeq\cos^{-1}(2v_{X}^{2}-1) \ .
\ee
In Fig.~\ref{figt:kin:1} we plot the opening angle as a function of the energy asymmetry for different values of the $X$ boson mass around $17$ MeV for all the transitions studied by ATOMKI. Note that the signal region for the $^8$Be case is all contained in $|y|\leq0.5$ in agreement with the ATOMKI experiment.
We then show in Fig.~\ref{fig:pdf_th} the normalized distribution of $\theta_\pm$, integrated over the asymmetry $y$, for the three nuclei considered by ATOMKI in the spin-0 boson hypotheses case, where 
the distribution only depends on phase space quantities. In the spin-1 case there is a dynamical dependence due to the polarization state of the $X$ boson which can modify this distribution, which is however expected to be qualitatively similar to the spin-0 case. As we see the opening angle distributions peak at the lowest end with peak values compatible with the ones reported by the experiments.
We further report in Tab.~\ref{tab:kin:3} in App.~\ref{sec:app_num_table} the predicted values for the 
opening angle at the signal peak for the various transitions investigated by ATOMKI.
Lastly, from Eq.~\eqref{thetapm} the invariant mass of the lepton pair reads
\be
m_{ee}^{2}=\frac{\omega^{2}}{2}\left[1-y^{2}+\delta^{2}- \cos\theta_{\pm} \sqrt{(1-\delta^{2}+y^{2})^{2}-4y^{2}}\right] \ ,
\ee
where here $\cos\theta_{\pm}$ is now a free parameter independent on the energy asymmetry.

%%%%%%%%%%%%%%%%%%%%%%%%%%%%%%%%%%%%%%%%%%%%%%%%%%%%%%%%%%%%%
%%%%%%%%%%%%%%%%%%%%%%%%%%%%%%%%%%%%%%%%%%%%%%%%%%%%%%%%%%%%%

\section{Signal computation: overview}
\label{sec:signal}

In this section we describe the multipole expansion formalism used in order to estimate the anomalous nuclear decay rates relevant for the ATOMKI experiment, see also~\cite{Kozaczuk:2016nma}.

\label{sec_sig}

\subsection{Nuclear states and processes}

\begin{figure}[t!]
\begin{center}
\includegraphics[width=0.3\textwidth]{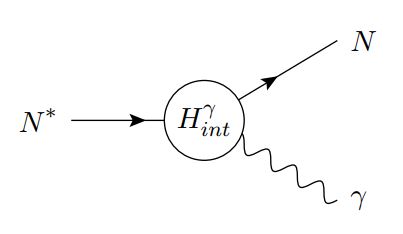}
\hskip 18pt
\includegraphics[width=0.3\textwidth]{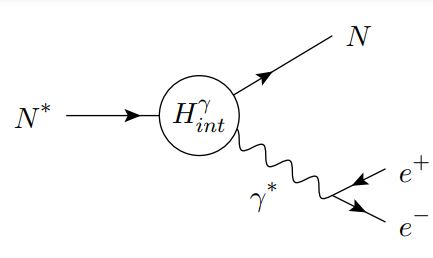}
\hskip 18pt
\includegraphics[width=0.3\textwidth]{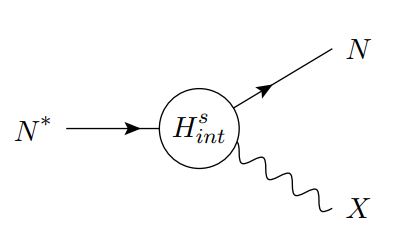}
\end{center}
\caption{Real $\gamma$-emission (left), IPC (center) and real $X$-emission (right) processes for the excited $N^*$ decay.}
\label{fig:diagrams}
\end{figure}

We describe the interaction of a spin $s$ $X$ boson to nuclear matter through the Hamiltonian
\be
H_{\rm int}^{s}=\begin{cases}
\int\!d^{3}\vec{r}\,\mathcal{S}\!(\vec{r}) X\!(\vec{r}) &\quad \text{if $s=0$ ,}\\
\int\!d^{3}\vec{r}\,\mathcal{J}_{\mu}\!(\vec{r}) X^{\mu}\!(\vec{r}) &\quad  \text{if $s=1$ ,}
\end{cases}
\ee
where the nuclear scalar density $\mathcal{S}$ and the nuclear current $\mathcal{J}^{\mu}=(\mathcal{J}^{0},\vec{\mathcal{J}})$ are quantum operators containing all the information of the matter fields. In the case of electromagnetic interaction the nuclear current ${\cal J}_\mu$ is replaced by the electromagnetic current ${\cal J}_\mu^\gamma$, which allows for $\gamma-$emission and IPC processes. At the lowest order in the interaction picture, the nuclear matrix element for the $X$ emission is given by
\be
\mathcal{T}_{fi}^{s}=
\bra{f,X}H_{\rm int}^{s}\ket{i}=\begin{cases}
\bra{f}\int\!d^{3}\vec{r}\,\mathcal{S}\!(\vec{r}) e^{-i\vec{k}\cdot\vec{r}} \ket{i} & \text{if $s=0$ ,}\\
\bra{f}\int\!d^{3}\vec{r}\,[\epsilon_{a}^{\mu}\!(\vec{k})]^{*}\mathcal{J}_{\mu}\!(\vec{r}) e^{-i\vec{k}\cdot\vec{r}} \ket{i} & \text{if $s=1$ ,}
\end{cases}
\ee
where $\vec k$ is the boson momentum, the index $a=0,\pm1$ labels the polarization of the vector boson and $\ket{i}$ and $\ket{f}$ indicate the nuclear matter initial and final states which are $\ket{f}=\ket{N;J_{f}M_{f}}$ and $\ket{i}=\ket{p+A;J_{p}M_{p};J_{A}M_{A};\vec{p}_{\rm CM}}$, where $\vec{p}_{\rm CM}$ is the proton momentum in the CM frame. We employ the narrow width approximation and factorize the excited resonance production from its decay, thereby assuming that the initial state is described by 
 $\ket{i_{*}}=\ket{N^{*};J_{*}M_{*}}$. In the following we want to compute the decay widths of the excited $N^*$ states for real $X$ emission in order to compare to the experimental results on this quantity reported by the ATOMKI collaboration. To make the calculation, it will turn out to be useful to expand the nuclear matrix elements in terms of spherical tensor operators, which will allow to also compute the electromagnetic real $\gamma$-emission and IPC processes, which we report in App.~\ref{app:sec_EM}. The diagrams for these three processes are shown in Fig.~\ref{fig:diagrams}.
To perform the calculation we will expand the nuclear matrix elements in terms of spherical tensor operators through a multipole expansion.

\subsection{Multipole expansion}

Spherical operators ${\cal O}_{JM}$ are irreducible tensor operators which satisfy the Wigner-Eckart theorem \cite{book:17167}
\be
\begin{split}\label{eq:wigner}
\braket{J_{f}M_{f}|O_{J,-M}|J_{i}M_{i}}=\frac{(-1)^{J_{i}-M_{i}}}{\sqrt{2J+1}}\braket{J_{f}M_{f};J_{i},-M_{i}|J_{f}J_{i};J,-M}\braket{J_{f}||O_{J}||J_{i}}.
\end{split}
\ee
The reduced matrix element $\braket{J_{f}||O_{J}||J_{i}}$ contains all the physical information of the operator while its behavior under rotation is completely set by the Clebsh-Gordan coefficient  $C^{J-M}_{J_f M_f J_i -M_i}$. We define the spherical 
operators\footnote{$Y_{JM}$, $j_{J}(x)$ and $\textbf{Y}_{J\ell M}(\hat{r})$ are respectively the 
spherical harmonics, the vector Bessel functions and the vector spherical harmonics, see \cite{book:17167}.} 
\begin{align}
\mathcal{G}_{JM} & =
\int d^{3}\vec{r}\,j_{J}(kr)Y_{JM}(\hat{r})\mathcal{S}(\vec{r}) \ ,\label{mso5} \\
\mathcal{M}_{JM} & =\int d^{3}\vec{r}\,j_{J}(kr)Y_{JM}(\hat{r})\mathcal{J}^{0}(\vec{r}) \ , \label{mso1}\\
\mathcal{L}_{JM}  &=\frac{i}{k}\int d^{3}\vec{r}\,\vec{\nabla}[j_{J}(kr)Y_{JM}(\hat{r})]\cdot\vec{\mathcal{J}}(\vec{r})\ , \label{mso2}\\
\mathcal{T}_{JM}^{\rm el} & =\frac{1}{k}\int d^{3}\vec{r}\,\vec{\nabla}\times[j_{J}(kr)\textbf{Y}_{JJM}(\hat{r})]\cdot\vec{\mathcal{J}}(\vec{r}) \ ,\label{mso3}\\
\mathcal{T}_{JM}^{\rm mag} & =\int d^{3}\vec{r}\,[j_{J}(kr)\textbf{Y}_{JJM}(\hat{r})]\cdot\vec{\mathcal{J}}(\vec{r}) \ , \label{mso4}
\end{align}
where $r = |\vec r|$ and expand the nuclear matrix elements as a sum of reduced matrix elements of spherical operators. In the case of interest of the emission of the $X$ boson in the process $N^* \to N + X$ one finds
\begin{align}
\mathcal{T}_{fi_{*}}^{s=0}  = & \sum_{\substack{J\geq 0,\\ |M|\leq J}}(-i)^{J}\sqrt{4\pi} C^{J-M}_{J_f M_f J_{*} -M_{*}} \braket{f||\mathcal{G}_{J}||i_{*}}D_{-M,0}^{(J)}(\phi,\theta,\beta) \ ,  \\ 
\mathcal{T}_{fi_{*}}^{s=1}=&\sum_{\substack{J\geq 0,\\ |M|\leq J}}(-i)^{J}\sqrt{4\pi}\delta_{a0} C^{J-M}_{J_f M_f J_{*} -M_{*}} \braket{f||\left[ \frac{k}{m}\mathcal{M}_{J}-\frac{\omega}{m}\mathcal{L}_{J}\right]||i_{*}}D_{-M,-a}^{(J)}(\phi,\theta,\beta) + \nonumber \\ 
-& \sum_{\substack{J\geq 1,\\ |M|\leq J,\\ \lambda=\pm1}}(-i)^{J}\sqrt{2\pi}\delta_{a\lambda} C^{J-M}_{J_f M_f J_{*} -M_{*}} \braket{f||\left[ \mathcal{T}_{J}^{el}+\lambda\mathcal{T}_{J}^{mag}\right]||i_{*}} D_{-M,-a}^{(J)}(\phi,\theta,\beta) \ ,
\end{align}
where here the indices $J$ and $M$ denote the total angular momentum of the emitted boson (sum of its spin and relative angular momentum with the $N$ nucleus) and its projection. The rotation $D$ matrices play the role of the wave function\footnote{See~\cite{book:17167} for the definition of the the $D$ functions.}, whose moduli squared give the probability for the $X$ boson to be emitted in the $(\phi,\theta)$ direction with $\beta$ defining a rotation along this direction\footnote{For the emission of a real boson, the angle $\beta$ is unphysical since it vanishes once the amplitudes are squared. It become physical in the IPC process.}. An explicit calculation for the unpolarized decays gives
\begin{align}
\Gamma_{X}^{s=0} & =\frac{2k}{2J_{*}+1}\Biggl\{ \sum_{J\geq 0}\left|\braket{f||\mathcal{G}_{J}||i_{*}}\right|^{2}\Biggr\} \ , \label{dl2}\\
\Gamma_{X}^{s=1} & =\frac{2k}{2J_{*}+1}\Biggl\{ \sum_{J\geq 0}\left|\braket{f||\left[ \frac{k}{m}\mathcal{M}_{J}-\frac{\omega}{m}\mathcal{L}_{J}\right]||i_{*}}\right|^{2}+ \sum_{J\geq 1}\left[\left|\braket{f|| \mathcal{T}_{J}^{el}||i_{*}}\right|^{2} + \left|\braket{f||\mathcal{T}_{J}^{mag}||i_{*}}\right|^{2}\right]\Biggr\} \ . \label{dl1}
\end{align}
In the case where the vector boson is coupled to a conserved current, {\emph{i.e.}} $\partial^\mu {\cal J}_\mu =0$, a simplification occurs.  By assuming the nuclear initial and final state to be eigenstates of the nuclear Hamiltonian
 the continuity equation $\vec{\nabla}\cdot\vec{\mathcal{J}}=-\frac{\partial\mathcal{J}^{0}}{\partial t}$
yields
\be\label{LtoM}
\omega\braket{f||\mathcal{M}_{J}||i_{*}}=k\braket{f||\mathcal{L}_{J}||i_{*}} \ ,
\ee
and the partial width for the emission of a vector $X$ boson then reduces to 
\begin{gather}
\Gamma_{X}^{s=1}=\frac{2k}{2J_{*}+1}\Biggl\{ \left(\frac{m}{k}\right)^{2}\sum_{J\geq 0}\left|\braket{f||\mathcal{M}_{J}||i_{*}}\right|^{2}+ \sum_{J\geq 1}\left[\left|\braket{f|| \mathcal{T}_{J}^{\rm el}||i_{*}}\right|^{2} + \left|\braket{f||\mathcal{T}_{J}^{\rm mag}||i_{*}}\right|^{2}\right]\Biggr\} \ . \label{dlcc}
\end{gather}
The above results are equally useful for the electromagnetic processes once we substitute the electromagnetic current in the spherical operators and put $m_{\gamma}=0$, {\emph{i.e.}} $k=\omega$.

\subsection*{Selection rules}

The angular momentum conservation law, encoded in the Clebsh-Gordan coefficient of Eq.~\eqref{eq:wigner}, states that the matrix element of the spherical operators vanishes unless the following conditions are satisfied
\begin{align}
& |J_{f}-J_{*}|\leq J\leq J_{f}+J_{*}\ , \nonumber \\
& M=M_{*}-M_{f} \ .
\end{align}
Moreover, if the $X$ boson has a definite parity $\pi_X$,  additional constraints on the matrix elements come from the requirement of parity conservation. By denoting the relative angular momentum between the boson and $N$ as $L$, one has
\begin{equation}
\pi_{*}=\pi_{f}\pi_{X}(-1)^{L} \ .
\end{equation}
We report in Tab.~\ref{tab:L} the relative angular momentum between the $X$ boson and $N$ in the various decay processes, based on the $S^\pi$ spin-parity assignments. One sees that  
a pure scalar solution to the $^8$Be anomaly is excluded,
while a pseudoscalar state can explain only the $^8$Be and $^4$He anomaly, if the latter is dominated by the $^4$He(21.01) excited state transition, but not the $^{12}$C one. On the other side a vector or axial-vector candidate can simultaneously explain all the three anomalies, but again only one of the two $^4$He resonant states can contribute to the signal process.

\begin{table}[t!]
\begin{center}
\begin{tabular}{c || cccc}
{\bf Process} &  \multicolumn{4}{c}{{\bf $X$ boson spin parity}}\\
$N^{*}\to N$ & $S^\pi =1^{-}$ & $S^\pi = 1^{+}$ & $S^\pi =0^{-}$ & $S^\pi =0^{+}$ \\
\midrule
\midrule
$^{8}$Be(18.15) $\to$ $^{8}$Be & 1 & 0, 2 & 1 & \slash \\
\midrule
$^{8}$Be(17.64) $\to$ $^{8}$Be & 1 & 0, 2 & 1 & \slash \\
\midrule\midrule
$^{4}$He(21.01) $\to$ $^{4}$He & \slash & 1 & 0 & \slash \\
\midrule
$^{4}$He(20.21) $\to$ $^{4}$He & 1 & \slash & \slash & 0 \\
\midrule\midrule
$^{12}$C(17.23) $\to$ $^{12}$C & 0, 2 & 1 & \slash & 1 \\
\end{tabular}
\end{center}
\caption{Relative angular momentum between the $X$ boson and $N$ in the various decays, based on its possible parity-spin assignments. Note that parity conservation prohibits a pure scalar solution to the Beryllium anomaly.}
\label{tab:L}
\end{table}

\subsection{Long wavelength approximation}

The nuclear radius is approximately given by $R \simeq R_0 A^{\frac{1}{3}} \simeq 6.1 \times 10^{-3} A^{\frac{1}{3}}\;$MeV$^{-1}$ \cite{book:80102}, which implies that in all the cases of interest the nucleus size is significantly smaller than the boson wavelength $k^{-1}\sim (10\;{\rm MeV})^{-1}$. We can thus expand the spherical Bessel function for small $k r$ as
\begin{equation}
\label{lwl}
j_{J}(kr)\simeq\frac{(kr)^{J}}{(2J+1)!!} \ ,
\end{equation}
with higher order corrections giving a contribution of order $(kr)^2\simeq 1\%$ with respect to the leading one for the cases of interest, which can therefore be neglected.
For the spherical operators $\mathcal{M}_{JM}$, $\mathcal{L}_{JM}$ and $\mathcal{G}_{JM}$ the expressions of Eq.~\eqref{mso5}, Eq.~\eqref{mso1} and Eq.~\eqref{mso2}  then read
\begin{align}
\label{lwl1}
& \mathcal{G}_{JM} \simeq\frac{k^{J}}{(2J+1)!!}\int d^{3}\vec{r}\,r^{J}Y_{JM}\mathcal{S}(\vec{r}) \ , \\
& \mathcal{M}_{JM} \simeq\frac{k^{J}}{(2J+1)!!}\int d^{3}\vec{r}\,r^{J}Y_{JM}\mathcal{J}^{0}(\vec{r}) ,\\
& \mathcal{L}_{JM} \simeq\frac{1}{i}\frac{k^{J-1}}{(2J+1)!!}\int d^{3}\vec{r}\,r^{J}Y_{JM}\vec{\nabla}\cdot\vec{\mathcal{J}}(\vec{r}) \ .
\end{align}
An exception occurs for the monopole case  $\mathcal{L}_{00}$, since it identically vanish at this order. The first contribution thus arises at the next order in the $kr$ expansion and is given by
\begin{equation}
\label{lwl2}
\mathcal{L}_{00}\simeq\frac{ik}{6}\int d^{3}\vec{r}\,r^{2}Y_{00}\vec{\nabla}\cdot\vec{\mathcal{J}}(\vec{r}) \ .
\end{equation}
The second order expansion is also needed for the ${\cal M}_{00}$ monopole expression in the case of a conserved current. This is due to the fact that in this case the integral over space of ${\cal J}^0(\vec r)$ defines the generator $Q$ of the symmetry associated with it. Then, with $| i \rangle$ and $| f \rangle$ orthogonal eigenstates of the Hamiltonian, one as  $\langle f|Q|i\rangle\propto\langle f|i\rangle=0$. It follows that the first contribution to the operator ${\cal M}_{00}$ is given by
\begin{equation}
\label{lwl3}
\mathcal{M}_{00}\simeq -\frac{k^{2}}{6}\int d^{3}\vec{r}\,r^{2}Y_{00}\mathcal{J}^{0}(\vec{r}) \ . 
\end{equation}
Let's now consider the operators $\mathcal{T}_{JM}^{\rm el}$ and $\mathcal{T}_{JM}^{\rm mag}$ of Eq.~\eqref{mso3} and Eq.~\eqref{mso4}. By using the identity 
\begin{equation}
\textbf{L}Y_{JM}=-i(\vec{r}\times\vec{\nabla})Y_{JM}=\sqrt{J(J+1)}\textbf{Y}_{JJM} \ ,
\end{equation}
they can be rewritten at the first order in the $kr$ expansion as~\cite{book:80102}
\begin{align}
\label{lwl4}
& \mathcal{T}_{JM}^{\rm mag} \simeq \frac{i k^{J}}{(2J+1)!!}\sqrt{\frac{J+1}{J}}\int d^{3}\vec{r}\left\{ \vec{\mu}(\vec{r}) + \frac{1}{J+1}\vec{r}\times\vec{J}_{irr}(\vec{r}) \right\}\cdot\vec{\nabla}(r^{J}Y_{JM}) , \\
& \mathcal{T}_{JM}^{\rm el}  \simeq \frac{1}{i}\frac{k^{J-1}}{(2J+1)!!}\sqrt{\frac{J+1}{J}}\int d^{3}\vec{r}\left\{ \vec{\nabla}\cdot\vec{J}_{\rm irr}(\vec{r}) + \frac{k^{2}}{J+1}\vec{\nabla}\cdot[\vec{r}\times\vec{\mu}(\vec{r})] \right\}r^{J}Y_{JM} \ ,
\end{align}
where the vector current has been split into an irrotational field $\vec{\cal J}_{\rm irr}$ and a solenoidal field $\vec{\nabla}\times\vec{\mu}$ as  $\vec {\cal J} = \vec{\cal J}_{\rm irr} + \vec{\nabla}\times\vec{\mu}$ in virtue of the Helmholtz's theorem.  For a conserved current,  the matrix element expression of $\mathcal{T}_{J}^{\rm el}$ can be simplified to
\begin{equation}
\label{lwl5}
\braket{ f|\mathcal{T}_{JM}^{\rm el}|i_{*}}\simeq\braket{f|\frac{k^{J}}{(2J+1)!!}\sqrt{\frac{J+1}{J}}\int d^{3}\vec{r}\left\{ \frac{\omega}{k}r^{J}Y_{JM}\mathcal{J}^{0}(\vec{r}) - \frac{ik}{J+1}\vec{\mu}(\vec{r})\cdot[\vec{r}\times\vec{\nabla}(r^{J}Y_{JM})] \right\}|i_{*}} \ ,
\end{equation}
by again using the continuity equation.

\subsection{Non relativistic expansion for nuclear operators} \label{nonrel}

\begin{table}[t!]

\begin{center}
\scalebox{0.94}{
\begin{tabular}{c||c|c|c}
 \multicolumn{4}{c}{{\bf{Spin-1 case}}}\vspace{0.3 cm} \\
											    & ${\cal J}^0(\vec r)$ & $\vec {\cal J}_{\rm irr}(\vec r)$ & $\vec \mu(\vec r)$ \\
\hline\hline
$C_{p}\bar{p}\gamma^{\mu}p+C_{n}\bar{n}\gamma^{\mu}n$ &  $\sum_{j=1}^{A}C_{j}\delta_{\vec r, \vec r_j}$ & $\sum_{j=1}^{A}\frac{C_{j}}{2m_{j}}\{\vec{p}_{j},\delta_{\vec r, \vec r_j}\}$& $\sum_{j=1}^{A}\frac{C_{j}}{2m_{j}}\vec{\sigma}_{j}\delta_{\vec r, \vec r_j}$ \\
\hline
$a_{p}\bar{p}\gamma^{\mu}\gamma^{5}p+a_{n}\bar{n}\gamma^{\mu}\gamma^{5}n$ & $\sum_{j=1}^{A}\frac{a_{j}}{2m_{j}}\{\vec{\sigma}_{j}\cdot\vec{p}_{j},\delta_{\vec r, \vec r_j}\}$ &  $\sum_{j=1}^{A}a_{j}\vec{\sigma}_{j}\delta_{\vec r, \vec r_j}$ & \slash \\
\hline
$\frac{\kappa_{p}}{2m_{p}}\partial_{\nu}(\bar{p}\sigma^{\mu\nu}p)+\frac{\kappa_{n}}{2m_{n}}\partial_{\nu}(\bar{n}\sigma^{\mu\nu}n)$ &\slash & \slash &  $\sum_{j=1}^{A}\frac{\kappa_{j}}{2m_{j}}\vec{\sigma}_{j}\delta_{\vec r, \vec r_j}$\\
\end{tabular}
}
\vskip 20pt
\scalebox{0.94}{
\begin{tabular}{c||c}
 \multicolumn{2}{c}{{\bf{Spin-0 case}}}\vspace{0.3 cm} \\
											    & ${\cal S}(\vec r)$   \\
\hline\hline
$z_{p}\bar{p}p+z_{n}\bar{n}n$ &  $\sum_{j=1}^{A}z_{j}\delta_{\vec r, \vec r_j}$   \\
\hline
$h_{p}i\bar{p}\gamma^{5}p+h_{n}i\bar{n}\gamma^{5}n$ &  $\sum_{j=1}^{A}\frac{h_{j}}{2m_{j}}\vec{\sigma}_{j}\cdot\vec{\nabla}[\delta_{\vec r, \vec r_j}]$   
\end{tabular}
}
\end{center}
\caption{Leading term of the non relativistic expansion for the relativistic vector current, the relativistic axial current and the anomalous magnetic moment terms (upper table) and for the scalar and pseudoscalar density (lower table). $\delta_{\vec r, \vec r_j}=\delta(\vec r - \vec r_j)$. }
\label{eq:non_rel_matching}
\end{table}

Through statistical considerations~\cite{book:14848}, the maximal kinetic energy $E_{\rm c}$ per nucleon in the nucleus is estimated to be around 30\;MeV, implying that 
a nucleus can be then modeled as a quantum mechanical system of non relativistic point-like nucleons~\cite{book:80102}. One can then take the non relativistic limit of the nuclear operator and write it in first quantization formalism. The nuclear operator is given by
\be
\mathcal{O}(\vec{r})=\sum_{i=1}^{A}\hat{\mathcal{O}}_{i}^{(1)}(\vec{r}-\vec{r}_{i}) \ ,
\label{eq:nuc_op}
\ee
with the single particle operator $\hat{\mathcal{O}}_{i}^{(1)}(\vec{r}-\vec{r}_{i})\propto\delta(\vec{r}-\vec{r}_{i})$ in the nucleon point-like approximation. We want to match the expression of Eq.~\eqref{eq:nuc_op} 
 with its relativistic counterpart, where the nucleons are described in terms of quantum fields $p(x)$ and $n(x)$ and the nucleon operators are bilinears in $p(x)$ and $n(x)$.
We report in Tab.~\ref{eq:non_rel_matching} the leading terms of the non relativistic expansion for the vector current, axial current, anomalous magnetic moment, scalar and pseudoscalar densities. For operators which are even or odd under parity, as it is in our case, higher order terms in the expansions are of order of $\frac{p_N^2}{m_N^2}\sim 6\times 10^{-2}$ with respect to the leading one, and can then be safely neglected.
For the specific case of the pseudoscalar density one also has that
\begin{equation}
\lim_{p'\rightarrow p}\bar{u}(p')\gamma^{5}u(p)=0 \ ,
\end{equation}
where $u$ are the spinors which enter the quantum field expression. Hence the non relativistic expansion of the pseudoscalar current only contains terms proportional to $\vec k = \vec p - \vec p^\prime$. Since for the effective non relativistic operator it holds the substitution~\cite{book:80102}
\begin{equation}
i\vec{\nabla}\rightarrow\vec{k}
\end{equation} 
the non relativistic expansion of the pseudoscalar density only contains operators given by a total divergence, meaning that the pseudoscalar density is a derivative coupling. The monopole operator $\mathcal{G}_{00}$ then vanishes at first order in the long-wavelength expansion and the leading contribution is then given by
\begin{equation}
\label{ggg}
\mathcal{G}_{00}\simeq-\frac{k^{2}}{6}\int d^{3}\vec{r}\,r^{2}Y_{00}\mathcal{S}(\vec{r}) \ .
\end{equation}
The techniques introduced in this section will be implemented in the next to derive the theoretical decay rates of the nuclear transitions.

\section{Signal computation: $X$ dynamics}
\label{sec_X17}

With the formalism described in the previous section we are now ready to describe the BSM dynamics of $e^+ e^-$ emission from the $X$ boson, illustrated in Fig.~\ref{fig:diagrams}. We refer to App.~\ref{app:sec_EM} for the details of the SM processes of real $\gamma$ emission and IPC. We parametrize the interaction of the $X$ boson with the scalar density $\mathcal{S}$ for the spin-0 case and the nuclear current $\mathcal{J}^{\mu}$ for the spin-1 cases in terms of effective couplings as\footnote{Effective nucleon operators are in principle also function of form factors $F(q^2)$. In all practical cases however the transferred momentum is much smaller than the hadron scale $\Lambda_{\rm QCD}$, so that we approximate the form factors as constants.}$^,$\footnote{We neglect electric dipole moment (EDM) operators since, by naive counting analysis, they would contribute at a higher order than the axial current $\bar \psi \gamma^\mu \gamma^5 \psi$. Moreover, they can only be generated by EDM effective quark operators, so that they will generally be suppressed by loop effects.}
\begin{align}
{\cal L}_{S^\pi = 0^+} & =  z_{p}\bar{p}p X+z_{n}\bar{n}n X \ , \\
{\cal L}_{S^\pi = 0^-} & = i h_{p}\bar{p}\gamma^{5}p X+i h_{n}\bar{n}\gamma^{5}n X \ , \\
 {\cal L}_{S^\pi = 1^-} & = C_{p}\bar{p}\gamma^{\mu}p X_\mu+C_{n}\bar{n}\gamma^{\mu}n X_\mu+\frac{\kappa_{p}}{2m_{p}}\partial_{\nu}(\bar{p}\sigma^{\mu\nu}p) X_\mu+\frac{\kappa_{n}}{2m_{n}}\partial_{\nu}(\bar{n}\sigma^{\mu\nu}n) X_\mu \ , \\
  {\cal L}_{S^\pi = 1^+} & =\ a_{p}\bar{p}\gamma^{\mu}\gamma^{5}p X_\mu+\ a_{n}\bar{n}\gamma^{\mu}\gamma^{5}n X_\mu \ ,
\end{align}
see App.~\ref{qton}.
Although the pure scalar hypothesis is not able to explain the anomaly observed in the $^8$Be decay, we'll present for completeness explicit expressions also in this case, since it can anyway affect the $^4$He and $^{12}$C decays, see Tab.~\ref{tab:L} and it can be relevant in the mixed parity hypothesis. The effective matching between these effective interactions and the UV interactions of the $X$ boson with quark and gluons are reported in App.~\ref{qton}.
Once the $X$ boson is produced from the nuclear collision, it decays to an $e^+e^-$ pair with a branching ratio which depends 
on the size of the $X$ coupling to electrons.

\subsection{Spherical operators}

In the non relativistic and long wavelength approximation the spherical operators with $J=0,1$ for the various $S^\pi$ assignment for the $X$ boson are given by
\subsubsection*{Scalar case $S^\pi = 0^+$}
For the $J=0$ term we go here beyond the leading order in the long wavelength and non relativistic expansion since leading order term proportional to the identity gives an identically vanishing contribution. One has\footnote{$\hat e_M = \sqrt{4\pi} Y_{1M}(\frac{\vec r}{r})$.}
\begin{align}
& \mathcal{G}_{00}\simeq \frac{1}{\sqrt{4\pi}}\sum_{s=1}^{A}z_{s}\left[ 1 - \frac{p_{s}^{2}}{2m_{\N}^{2}} - \frac{k^{2}r_{s}^{2}}{6} \right] , \\
&\mathcal{G}_{1M}\simeq \frac{k}{3}\sqrt{\frac{3}{4\pi}}\sum_{s=1}^{A}z_{s}\vec{r}_{s}\cdot \hat{e}_{M} ,
\end{align}
where $z_{s}=z_{p}$ ($z_{s}=z_{p}$) if the $s$-th nucleon is a proton (neutron). A similar notation is adopted for the rest of the section.

\subsubsection*{Pseudoscalar case $S^\pi = 0^-$}
For the pseudoscalar case the relevant spherical operators are
\begin{align}
& \mathcal{G}_{00}\simeq \frac{k^{2}}{6m_{\N}}\frac{1}{\sqrt{4\pi}}\sum_{s=1}^{A}h_{s}(\vec{r}_{s}\cdot\vec{\sigma}_{s})\equiv \frac{k^{2}}{12m_{\N}}\frac{1}{\sqrt{4\pi}}[(h_{p}+h_{n})\hat{d}_{0}^{\sigma}+(h_{p}-h_{n})\hat{d}_{3}^{\sigma}] \ , \\
& \mathcal{G}_{1M}\simeq -\frac{k}{6m_{\N}}\sqrt{\frac{3}{4\pi}}\sum_{s=1}^{A}h_{s}\vec{\sigma}_{s}\cdot \hat{e}_{M}\equiv -\frac{k}{6m_{\N}}\sqrt{\frac{3}{4\pi}}[h_{p}\hat{\sigma}_{M}^{(p)}+h_{n}\hat{\sigma}_{M}^{(n)}] \ ,
\end{align}
where for ${\cal G}_{00}$ we have split the expression among the isoscalar and isovector contributions.

\subsubsection*{Vector case $S^\pi = 1^-$}

In this case conservation of the vector current implies a relation between the operator ${\cal L}_{JM}$ and ${\cal M}_{JM}$, {\emph{c.f.r.}} Eq.~\eqref{LtoM}, and one has

\begin{align}
& \mathcal{M}_{00}\simeq -\frac{k^{2}}{6}\frac{1}{\sqrt{4\pi}}\sum_{s=1}^{A}C_{s}r_{s}^{2}\equiv-\frac{ek^{2}}{6}\rho^{(X)}\ , \label{monopole}\\
&\mathcal{M}_{1M}\simeq \frac{k}{3}\sqrt{\frac{3}{4\pi}}\sum_{s=1}^{A}C_{s}\vec{r}_{s}\cdot \hat{e}_{M}\equiv \frac{ek}{3}d_{M}^{(X)}\ , \\
&\mathcal{T}_{\rm 1M}^{el}\simeq \frac{\sqrt{2}\omega}{3}\sqrt{\frac{3}{4\pi}}\sum_{s=1}^{A}C_{s}\vec{r}_{s}\cdot \hat{e}_{M}\equiv \frac{\sqrt{2}e\omega}{3}d_{M}^{(X)}\ , \\
&\mathcal{T}_{1M}^{\rm mag}\simeq \frac{i\sqrt{2}k}{3}\frac{1}{2m_{\N}}\sqrt{\frac{3}{4\pi}}\sum_{s=1}^{A}[C_{s}(\vec{r}_{s}\times\vec{p}_{s})+(C_{s}+\kappa_{s})\vec{\sigma}_{s}]\cdot \hat{e}_{M}\equiv \frac{i\sqrt{2}k\mu_{\N}}{3}\mu_{M}^{(X)} \ .
\end{align}

\subsubsection*{Axial vector case $S^\pi = 1^+$}

Finally the spherical operators for the axial vector case read
\begin{align}
& \mathcal{M}_{00}\simeq \mathcal{M}_{1M} \simeq 0 \, , \\
& \mathcal{L}_{00}\simeq -\frac{ik}{3}\frac{1}{\sqrt{4\pi}}\sum_{s=1}^{A}a_{s}(\vec{r}_{s}\cdot\vec{\sigma}_{s})\equiv -\frac{ik}{6}\frac{1}{\sqrt{4\pi}}[(a_{p}+a_{n})\hat{d}_{0}^{\sigma}+(a_{p}-a_{n})\hat{d}_{3}^{\sigma}] \, , \\
& \mathcal{L}_{1M}\simeq \frac{i}{3}\sqrt{\frac{3}{4\pi}}\sum_{s=1}^{A}a_{s}\vec{\sigma}_{s}\cdot \hat{e}_{M}\equiv\frac{i}{3}\sqrt{\frac{3}{4\pi}}[a_{p}\hat{\sigma}_{M}^{(p)}+a_{n}\hat{\sigma}_{M}^{(n)}] \, , \\
& \mathcal{T}_{1M}^{\rm el}\simeq \frac{i\sqrt{2}}{3}\sqrt{\frac{3}{4\pi}}\sum_{s=1}^{A}a_{s}\vec{\sigma}_{s}\cdot \hat{e}_{M}\equiv\frac{i\sqrt{2}}{3}\sqrt{\frac{3}{4\pi}}[a_{p}\hat{\sigma}_{M}^{(p)}+a_{n}\hat{\sigma}_{M}^{(n)}] \, , \\
&  \mathcal{T}_{1M}^{\rm mag}\simeq \frac{ik}{3\sqrt{2}}\sqrt{\frac{3}{4\pi}}\sum_{s=1}^{A}a_{s}(\vec{r}_{s}\times\vec{\sigma}_{s})\cdot \hat{e}_{M}\equiv \frac{ik}{6\sqrt{2}}\sqrt{\frac{3}{4\pi}}[(a_{p}+a_{n})\hat{D}_{0M}^{\sigma}+(a_{p}-a_{n})\hat{D}_{3M}^{\sigma}] \, .
\end{align}

\subsection{Decay rates}\label{sec:nuclear_ME}

\begin{table}[t!]
\scalebox{0.82}{
\begin{tabular}{c||c|c|c}
 & {\bf $^8$Be} & {\bf $^4$He} & {\bf $^{12}$C} \\
 \hline \hline
\multirow{2}{*}{{\bf $0^+$}} & \multirow{2}{*}{\slash} 			      & {\bf $^4$He(20.21)}  						&\multirow{2}{*}{$\frac{2k^{3}}{27}(z_{p}-z_{n})^{2}|\braket{||d^{(\gamma)}||}|^{2}$}   \\
					     &  							      & $2k(z_{p}+z_{n})^{2}\left| \frac{k^{2}}{6e}\braket{\rho^{(\gamma)}} + \frac{1}{2m_{\N}}\braket{\hat{K}} \right|^{2}$ & \\
\hline		
\multirow{2}{*}{{\bf $0^-$}} & 	\multirow{2}{*}{$\frac{k^{3}}{72\pi m_{N}^{2}}|\braket{h_{p}\hat{\sigma}^{(p)} +h_{n} \hat{\sigma}^{(n)}}|^{2}$} 	 & 	{\bf$^4$He(21.01)}  & \multirow{2}{*}{\slash} \\
					     &						 &     $\frac{k^{5}}{228\pi m_{N}^{2}}(h_{p}+h_{n})^{2}|\braket{\hat{d}_{0}^{\sigma}}|^{2}$			& \\
\hline	
\multirow{2}{*}{{\bf $1^-$}} & \multirow{2}{*}{$\frac{4\mu_{\N}^{2}k^{3}}{27}|\braket{\mu^{(X)}}|^{2}$} 							& {\bf$^4$He(20.21)}& \multirow{2}{*}{$\frac{16\pi\alpha k\omega^{2}}{27}\left( 1+\frac{m^{2}}{2\omega^{2}} \right)|\braket{d^{(X)}}|^{2}$} \\
				  &   & 	$\frac{m^{2}k^{3}\alpha}{18}|\braket{\rho^{(X)}}|^{2}$ & \\ 
\hline				  
\multirow{2}{*}{{\bf $1^+$}} & \multirow{2}{*}{$\frac{k}{18\pi}\left(2+\frac{\omega^{2}}{m^{2}}\right)|\braket{a_{p}\hat{\sigma}^{(p)} +a_{n} \hat{\sigma}^{(n)}}|^{2}$} &{\bf$^4$He(21.01)} & \multirow{2}{*}{$\frac{k^{3}}{144\pi}(a_{p}-a_{n})^{2}|\braket{\hat{D}_{3}^{\sigma}}|^{2}$}\\
				   &  & $\frac{\omega^{2}k^{3}}{72\pi m^{2}}(a_{p}+a_{n})^{2}|\braket{\hat{d}_{0}^{\sigma}}|^{2}$ & 				  
\end{tabular}
}
\caption{Decay rates for the $^8$Be, $^4$He and $^{12}$C nuclear processes for the various spin assignment of the $X$ boson. In the case of the Helium transition for each spin-parity possibility we indicate the $^4$He excited state involved. In the expressions $\langle {\cal O} \rangle$ represents the matrix element between the ground state and the excited nucleus of the corresponding operator, {\emph{e.g.}} for the $^{12}$C transition in the $0^+$ case $\langle || d^{(\gamma)}||\rangle = \langle ^{12}{\rm C}|| d^{(\gamma)}||^{12}{\rm C}(17.23)\rangle$. For the $^4$He and $^{12}$C cases we only report the non vanishing isoscalar and isovector contributions respectively.}
\label{tab:rates}
\end{table}

We can now express the decay rates for the various spin-parity assignment of the $X$ boson in the case of the $^8$Be, $^4$He and $^{12}$C transitions. We report them Tab.~\ref{tab:rates}, expressed in function of nuclear matrix element of the relevant operators involved in the transition. Symmetry consideration allow to express these matrix elements in function of known ones. We list here the relevant relations

\subsubsection*{Beryllium matrix elements}
Assuming the static quark model $\kappa_{p}\simeq-\kappa_{n}\simeq 2(C_{p}-C_{n})$, see App. \ref{qton} for the details. From isospin symmetry then one has
\begin{gather}
\braket{^{8}\text{Be}||\mu^{(X)}||^{8}\text{Be}(17.64)}=\alpha_{1}\left(\frac{C_{p}+C_{n}}{e}\right)M1_{I=0}^{\gamma}+(\beta_{1}+\alpha_{1}\xi)\left(\frac{C_{p}-C_{n}}{e}\right)M1_{I=1}^{\gamma}\ ,\\
\braket{^{8}\text{Be}||\mu^{(X)}||^{8}\text{Be}(18.15)}=(-\alpha_{1}+\beta_{1}\xi)\left(\frac{C_{p}-C_{n}}{e}\right)M1_{I=1}^{\gamma}+\beta_{1}\left(\frac{C_{p}+C_{n}}{e}\right)M1_{I=0}^{\gamma}\ ,
\end{gather}
where the corresponding values are reported in App.~\ref{sec:em_ME}, while
we take from~\cite{Kozaczuk:2016nma}
\begin{align}
& \braket{\text{Be}||\hat{\sigma}^{(p)}||^{8}\text{Be}(18.15)}=-0.047(29)\ ,\,\,\,\,& \braket{\text{Be}||\hat{\sigma}^{(n)}||^{8}\text{Be}(18.15)}=-0.132(33)\ , \nonumber \\
& \braket{\text{Be}||\hat{\sigma}^{(p)}||^{8}\text{Be}(17.64)}=0.102(28) \ ,\,\,\,\,& \braket{\text{Be}||\hat{\sigma}^{(n)}||^{8}\text{Be}(17.64)}=-0.073(29)\ .
\end{align}

\subsubsection*{Helium matrix elements}

Isospin symmetry allows to relate
\be
\braket{^{4}\text{He}||\rho^{(X)}||^{4}\text{He}(20.21)}=\left(\frac{C_{p}+C_{n}}{e}\right)\braket{^{4}\text{He}||\rho^{(\gamma)}||^{4}\text{He}(20.21)}  \ ,
\ee
where again the electromagnetic matrix element is reported in 
App.~\ref{sec:em_ME} while we take from~\cite{Horiuchi:2009zza}
\begin{equation}
|\braket{^{4}\text{He}||\hat{d}_{0}^{\sigma}||^{4}\text{He}(21.01)}|^{2}\simeq 15.5\,\text{fm}^{2}\simeq4\times10^{-4}\,\text{MeV}^{-2}
\end{equation}
but no uncertainty has been given. We will arbitrarily assume a $10\%$ error on the matrix element in our calculation. Finally, to the best of our knowledge, the matrix element of the operator $\hat{K} = \sum_{s} p_{s}^{2}/2m_s$ has never been evaluated so far. 

\subsubsection*{Carbon matrix elements}

Isospin symmetry allows to relate
\be
\braket{^{12}\text{C}||d^{(X)}||^{12}\text{C}(17.23)}=\left(\frac{C_{p}-C_{n}}{e}\right)\braket{^{12}\text{C}||d^{(\gamma)}||^{12}\text{C}(17.23)} \ ,
\ee
whose values is reported in App.~\ref{sec:em_ME}, while the axial matrix element $\hat D_3^\sigma$ has not been evaluated, to the best of our knowledge.

\section{Results}\label{sec:res}

We present in this section our main results, deriving the possible range of the nucleon couplings to the $X$ particle that can explain both the $^8$Be and $^4$He anomalies, further commenting on the possibility of simultaneously explain the $^{12}$C one. We analyze all the scenarios where the $X$ boson has a definite parity, which implies that the pure scalar boson case is ruled out since it cannot explain the $^8$Be anomaly. The best fit value for the anomalous decay rate for the $^8$Be transition is~\cite{Krasznahorkay:2018snd} 
\be
\label{BeBestFit}
\frac{\Gamma({^8 \text{Be}}(18.15) \to {^8 \text{Be}} \, + \, X)}{\Gamma({^8 \text{Be}}(18.15) \to {^8 \text{Be}} \, + \, \gamma)}\ \text{BR}(X\to e^{+} e^{-})=(6\pm1)\times10^{-6}.
\ee
The ATOMKI collaboration observed no anomalous signal in the $^8$Be(17.64) transition in the first experiment~\cite{Krasznahorkay:2015iga} but later they reported a non vanishing best fit for this anomalous decay rate in a contribution to the proceedings of {\it International Symposium Advances in Dark Matter and Particle Physics 2016}~\cite{Krasznahorkay:2017gwn}. In the following we will consider only the $^8$Be(18.15) anomalous decay and we present the results with both the $^8$Be transitions in App. \ref{appBev}.
For the case of the $^4$He transition the total cross section is given by the sum of the two states populated in the experiment
\begin{equation}
\label{24}
\frac{\sigma_X}{\sigma_{E0}}=\frac{\Gamma({^4 \text{He}}(20.21) \to {^4 \text{He}} \, + \, X)}{\Gamma({^4 \text{He}}(20.21) \to {^4 \text{He}} \, + \, e^{+}e^{-})}\,+\,\frac{\sigma_{-}\Gamma_{+}}{\sigma_{+}\Gamma_{-}}\frac{\Gamma({^4 \text{He}}(21.01) \to {^4 \text{He}} \, + \, X)}{\Gamma({^4 \text{He}}(20.21) \to {^4 \text{He}} \, + \, e^{+}e^{-})} \ ,
\end{equation}
where $\Gamma_{\pm}$ is the total width of the $0^{\pm}$ excited state of Helium nucleus and 
\begin{gather}
\sigma_{+}=\sigma(p+{^{3}\text{H}}\to {^4 \text{He}}(20.21)), \qquad \sigma_{-}=\sigma(p+{^{3}\text{H}}\to {^4 \text{He}}(21.01)) \ .
\end{gather}
The ratio $\sigma_{-}/\sigma_{+}$ can be evaluate by the relation of Eq.~\eqref{resprod} in App.~\ref{NWAApp} in the narrow width approximation. The ATOMKI collaboration reported $\sigma_X=0.2 \ \sigma_{E0}$ \footnote{Differently from this work, the Authors of~\cite{Feng:2020mbt} took as experimental input the ratio of decay rates calculated in the experimental paper \cite{Krasznahorkay:2019lyl}.}, while no uncertainty is associated with this measurement. We then arbitrarily associate a relative error to the Helium best fit equal to the one from Beryllium measurement of Eq.~\eqref{BeBestFit}. Different spin parity assignments contribute to the rates of the two $^4$He excited states, see Tab.~\ref{tab:L}. 
If the  $X$ boson is a vector or a scalar state one has
\be\label{eq:He1}
\frac{\Gamma({^4 \text{He}}(20.21) \to {^4 \text{He}} \, + \, X)}{\Gamma({^4 \text{He}}(20.21) \to {^4 \text{He}} \, + \, e^{+}e^{-})}\ \text{BR}(X\to e^{+} e^{-})=0.20\pm0.03 \ ,
\ee
while if it's a pseudoscalar or an axial vector the best fit is
\be\label{eq:He2}
\frac{\Gamma({^4 \text{He}}(21.01) \to {^4 \text{He}} \, + \, X)}{\Gamma({^4 \text{He}}(20.21) \to {^4 \text{He}} \, + \, e^{+}e^{-})}\ \text{BR}(X\to e^{+} e^{-})= 0.87\pm0.14 \ ,
\ee
with $\Gamma({^4 \text{He}}(20.21) \to {^4 \text{He}} \, + \, e^{+}e^{-})=(3.3\pm1.0)\times10^{-4}$ eV \cite{Walcher:1970vkv}.
For the case of the $^{12}$C transition the recent results~\cite{Krasznahorkay:2022pxs} find the derived branching ratio for $X$ emission with respect to the $\gamma$ one to be $\sim 3.6(3)\times 10^{-6}$, {\emph{i.e.}}
\be
\label{CBestFit}
\frac{\Gamma({^{12}\text{C}}(17.23)\to{^{12}\text{C}}+X)}{\Gamma({^{12}\text{C}}(17.23)\to{^{12}\text{C}}+\gamma)}\ \text{BR}(X\to e^{+} e^{-})=3.6(3)\times 10^{-6}.
\ee

We now present our findings for the regions in the effective nucleon couplings parameter space for the various spin-parity assignments for the $X$ boson. 
In presenting our results we assume, for simplicity, $\text{BR}(X\to e^{+}e^{-})=1$. For different BR assumptions the derived allowed space in the nucleon effective couplings will be rescaled according to Eq.~\eqref{BeBestFit}, Eq.~\eqref{eq:He1} and Eq.~\eqref{eq:He2}.
We stress that our analysis relies on various assumptions, as the hypothesis of narrow width approximation for the nuclear production of the excited states. Other potential contribution, as for example direct capture processes, could potentially change the conclusions of our analysis.

\subsection{Pseudoscalar and mixed parity scenario}

\begin{figure}[t!]
\centering
{\includegraphics[width=.55\textwidth]{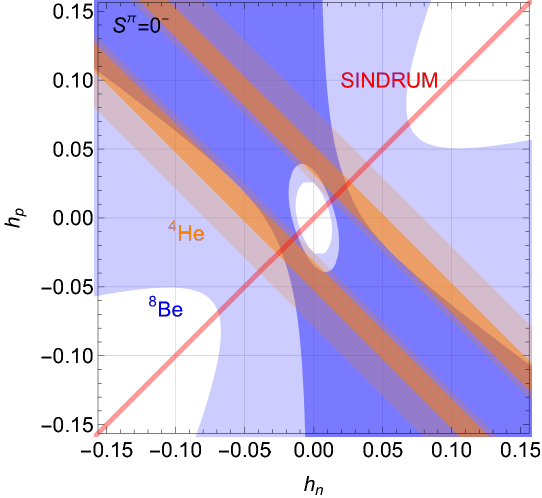}}
\caption{Regions of the $h_{n,p}$ effective nuclear couplings of a pure pseudoscalar states where the $^{8}$Be (blue) and $^{4}$He (orange) anomalous ATOMKI transition can be explained at $1\sigma$ or $2\sigma$. Inside the red region the SINDRUM bound is satisfied.}
\label{plot:pseudo}
\end{figure}

We summarize the results for the pure pseudoscalar scenario in Fig.~\ref{plot:pseudo}, where the shaded blue and orange areas represent the $1\sigma$ and $2\sigma$ compatibility regions with the ATOMKI $^8$Be and $^4$He anomalies respectively. We also overlay in red the region of parameter space satisfying the SINDRUM bound from $\pi^+ \to e^+ \nu_e X$ decay~\cite{SINDRUM:1986klz,Alves:2017avw}.
This is given in term of the pseudoscalar-pion mixing angle, linked to the isovector nucleon coupling as
\be
\theta_{X\pi} = \frac{f_\pi (h_p - h_n)}{2 g_A m_{p,n}}  \ ,
\ee
where $g_A \sim 1.27$ is axial nucleon factor and $f_\pi \sim 93\;$MeV is the pion decay constant, and reads
\be\label{eq:sindrum}
|\theta_{X\pi}|\lesssim \frac{10^{-4}}{\sqrt{{\rm BR}(X\to e^+ e^-)}} \ .
\ee
All together we see that the $^8$Be and $^4$He anomalies can be simultaneously satisfied for a range of effective nuclear coupling $h_{n,p}$ of ${\cal O}(10^{-2})$. 
However, the recent observation of an anomalous signal in the $^{12}$C transition would, if confirmed, exclude by itself the pure pseudoscalar scenario, see again Tab.~\ref{tab:L}. 
It's then interesting to entertain the possibility that the scalar $X$ boson has both scalar and pseudoscalar couplings. As already mentioned, because of parity conservation the scalar contribution to the $^8$Be transitions vanishes, so the latter processes only set a constraint on the range of the pseudoscalar couplings, which as we have shown in Fig.~\ref{plot:pseudo} are required to be of order of ${\cal O}(10^{-2})$. On the other side the $^4$He decays acquire a contribution from both the spin-parity state, although related to different nuclear resonances, see Tab.~\ref{tab:L}. As discussed in Sec.~\ref{sec:nuclear_ME} the value of the matrix element $\braket{^{4}\text{He}||\hat{K}||^{4}\text{He}(20.21)}$ is unknown. By neglecting its contribution one finds that the pure scalar contribution is dominant over the pseudoscalar one with a similar value for nucleon couplings $z_{p,n}~h_{p,n}\simeq 10^{-2}$, {\it i.e.}
\begin{align}
& \Gamma({^{4}\text{He}}(20.21)\to{^{4}\text{He}}+X)\simeq   \;6\times10^{-4}\text{eV}\,\left(\frac{z_{p}+z_{n}}{10^{-2}}\right)^{2} \ , \\
& \Gamma({^{4}\text{He}}(21.01)\to{^{4}\text{He}}+X)\simeq  \; 9.7\times10^{-6}\text{eV}\,\left(\frac{h_{p}+h_{n}}{10^{-2}}\right)^{2},
\end{align}
so that the theoretical predictions for the $^4$He transition is too large to match the ATOMKI results. We expect that this assertion holds even once $\braket{^{4}\text{He}||\hat{K}||^{4}\text{He}(20.21)}$ is also included. Hence, one is forced to conclude that the scalar isoscalar coupling $z_{p}+z_{n}$ is suppressed, at least respect the pseudoscalar one, leaving us with almost the same configurations as the pure pseudoscalar one of Fig.~\ref{plot:pseudo}. This conclusion is in agreement with earlier results~\cite{Feng:2020mbt}. On the other side for the $^{12}$C transition, the scalar isovector coupling would give
\be
\frac{\Gamma({^{12}\text{C}}(17.23)\to{^{12}\text{C}}+X)}{\Gamma({^{12}\text{C}}(17.23)\to{^{12}\text{C}}+\gamma)}\simeq 2.4\times10^{-6}\left(\frac{z_{p}-z_{n}}{10^{-2}}\right)^{2},
\ee
in agreement with the order of magnitude of the ATOMKI fit \eqref{CBestFit} if $z_{p}-z_{n}\simeq h_{p,n}\simeq\mathcal{O}(10^{-2})$ and all the three anomalous measurements can be simultaneously satisfied.

\subsection{Vector and axial vector scenarios}

\begin{figure}[t!]
\centering
{\includegraphics[width=.49\textwidth]{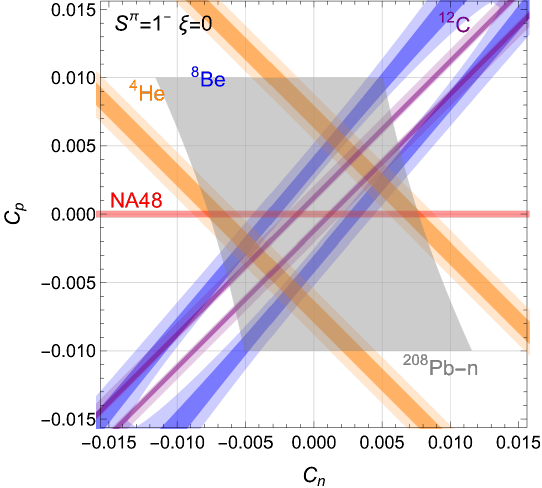}}\hfill
{\includegraphics[width=.49\textwidth]{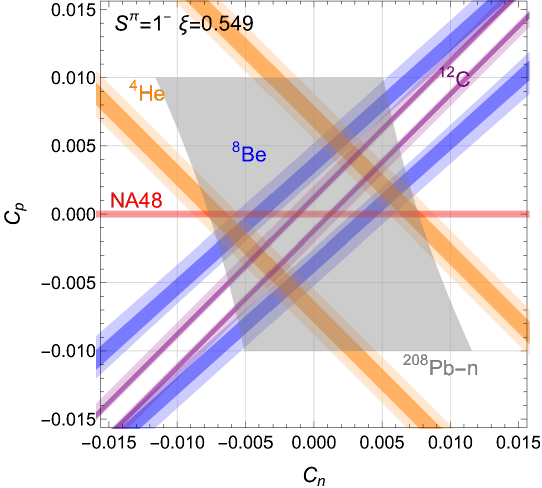}}\\
\vskip 6pt
{\includegraphics[width=.496\textwidth]{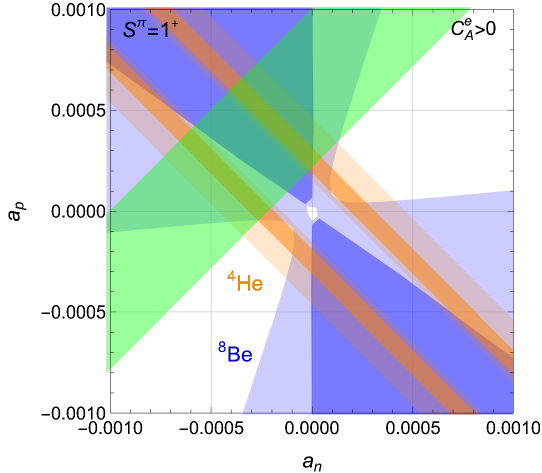}} \hfill
{\includegraphics[width=.496\textwidth]{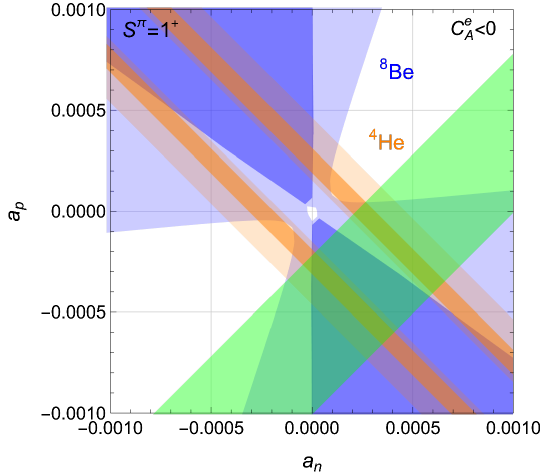}} 
\caption{{\it Upper panels:} Regions of the $C_{n,p}$ effective nuclear couplings of a pure vector state where the $^{8}$Be (blue), $^{4}$He (orange) and $^{12}$C (purple) anomalous ATOMKI transition can be explained at $1\sigma$ or $2\sigma$. Inside the red and the gray region, respectively, the NA48 and the $^{208}$Pb-n scattering bound are satisfied. In the left and right panel we assume $\xi=0$ and $\xi=0.549$ respectively, see App.~\ref{sec:em_ME} for details.
{\it Lower panels:}  Regions of the $a_{n,p}$ effective nuclear couplings of a pure axial vector state where the $^{8}$Be (blue) and $^{4}$He (orange) anomalous ATOMKI transition can be explained at $1\sigma$ or $2\sigma$. In the green region the KTeV anomaly in $\pi \to e^+ e^-$ decay can be satisfied, by assuming a positive (left panel) and negative (right panel) value for the $C_A^e$ axial coupling of the $X$ boson to electrons that can explain the anomalous $(g-2)_e$, see main text for more details.}
\label{plot:vector}
\end{figure}

We summarize the results for the spin-1 cases in Fig.~\ref{plot:vector},
with the same color code as Fig.~\ref{plot:pseudo} for the regions satisfying the ATOMKI anomalies.  In the upper panels we show the results for the $S^\pi = 1^-$ assignment for the $X$ boson. Here in the left and right plot we assume $\xi = 0$ and $\xi=0.549$ respectively, where $\xi$ represents non perturbative contribution to isospin breaking effects in the $^8$Be case, see~\cite{Feng:2016ysn} and App.~\ref{sec:em_ME} for details. For the vector case the strongest bound comes from the non observation from the NA48 experiment of the $\pi^0 \to \gamma X$ decay in dark photon searches~\cite{Batley:2015lha}.
This process receives two different contributions. One from the axial anomaly and a non anomalous one. The non anomalous contributions is proportional to the small quark masses~\cite{Sutherland:1967vf,10.2307/2415932} and can be neglected, while the anomalous one is proportional to the anomaly trace factor. One gets the bound
\be
|C_p| \times \sqrt{{\rm BR}(X\to e^+ e^-)} \lesssim 2.5 \times 10^{-4} \ ,
\ee
which implies a {\emph{protophobic}} nature for the $X$ boson. We show in Fig.~\ref{plot:vector} in red the region of parameter space where the {\emph{protophobia}} constraint is satisfied.
Another relevant bound comes from observations of the angular dependence of neutron-lead scattering. The exchange of new, weakly-coupled boson produces a Yukawa potential\footnote{The non-relativistic limit of the vector bilinears is spin independent, see Tab.~\ref{eq:non_rel_matching}, so the nucleon contributions of a nuclear state are added coherently, while they are spin dependent and are added incoherently for the axial and pseudoscalar cases. We then expect negligible contributions for the $S^\pi=0^-, 1^+$ cases, especially when considering a nucleus with null spin like $^{208}$Pb.} acting on the neutron, whose contribution has been constrained for the $^{208}$Pb-n scattering as \cite{Barbieri:1975xy}
\be
|C_{n}|\left|\frac{126}{208}C_{n}+\frac{82}{208}C_{p}\right|\lesssim 3.6\times10^{-5} \ .
\ee
We show in Fig.~\ref{plot:vector} in gray the region of parameter space where the $^{208}$Pb-n scattering constraint is satisfied. As it can be seen for the $\xi=0.549$ assignment, a combined explanation of the $^8$Be, blue region, and $^4$He, orange region, anomalies at $1\sigma$ is in tension with the NA48 constraint,
while it is possible at the $2\sigma$ level. Regarding the $^{12}$C ATOMKI anomaly in the case of a $S^\pi = 1^-$ state the relevant matrix element is known. In this case one then gets
\be
\frac{\Gamma({^{12}\text{C}}(17.23)\to{^{12}\text{C}}+X)}{\Gamma({^{12}\text{C}}(17.23)\to{^{12}\text{C}}+\gamma)}\simeq 2.64\times10^{-6}\left(\frac{C_{p}-C_{n}}{10^{-3}}\right)^{2},
\ee
in agreement with the order of magnitude of the ATOMKI fit \eqref{CBestFit} if $C_{p}-C_{n}\simeq\mathcal{O}(10^{-3})$. 
The $1\sigma$ and $2\sigma$ bands related to the $^{12}$C transition are shown in purple in the upper panels of Fig.~\ref{plot:vector}.
Note that, if confirmed, the $^{12}$C ATOMKI anomaly is in tension with a combined explanation of the $^8$Be and $^4$He anomalies and the {\emph{protophobia}} constraint.

On the other side an axial vector $S^\pi = 1^+$ state can explain both the $^8$Be and $^4$He ATOMKI anomalies, as shown in the lower panels of Fig.~\ref{plot:vector}, with axial couplings to the nucleon of ${\cal O}(10^{-4})$. Within the green shaded area the KTeV anomaly in $\pi^0 \to e^+ e^-$ decay can be explained for positive and negative values for the axial $X$ coupling to electrons $C_A^e$, see Sec.~\ref{sec:ktev} for details.
As regarding the possibility of also explaining the $^{12}$C ATOMKI anomaly the relevant nuclear matrix element, see Tab.~\ref{tab:rates}, is currently unknown. While no definite claim can be made until it becomes available, we can make an order of magnitude estimate on the size of the $\hat D_3^\sigma$  and speculate on the possibility of a combined explanation of all the three ATOMKI anomalies with an axial vector state.
We can expect that the isovector spin dipole would be of the order of the nuclear radius times the number of nucleons inside the nucleus. We can then estimate
\begin{equation}
\braket{^{12}\text{C}||\hat{D}_{3}^{\sigma}||^{12}\text{C}(17.23)}\simeq A\times R\simeq12\times2.75\,\text{fm}\simeq1.7\times10^{-1}\,\text{MeV}^{-1}.
\end{equation}
For a range of nucleon parameters $a_{p,n}\simeq {\cal O}(10^{-4})$, as suggested by Fig.~\ref{plot:vector}, one get an estimate for the anomalous $^{12}$C transition mediated by an axial $X$ boson of 
\begin{equation}
\frac{\Gamma({^{12} \text{C}}(17.23) \to {^{12} \text{C}} \, + \, X)}{\Gamma({^{12} \text{C}}(17.23) \to {^{12} \text{C}} \, + \, \gamma)}\simeq\mathcal{O}(10^{-6}) \ ,
\end{equation}
which is in order of magnitude accord with the ATOMKI result which predicts a value of $3.6(3)\times 10^{-6}$ for this rate~\cite{Krasznahorkay:2022pxs}. We stress again that this conclusion strongly depends on our order of magnitude estimate of the $\hat D_3^\sigma$ matrix element, which seems to indicate that an axial vector state might be favored for a combined explanation. However to properly test its consistency with the $^{12}$C anomalous transition, the relevant matrix element must be properly computed. Until then no definite conclusions can be drawn.
In a general scenario where both vector and axial couplings to nucleons are present, the decay width for the real $X$ emission is the direct sum of the two contributions. Assuming vector and axial couplings of the same order of magnitude, the axial contribution would typically dominate over the vector one.

Intriguingly, for the case a pure axial boson $S^\pi = 1^+$, in the parameter space where the $^4$He and $^8$Be anomalies can be explained, other experimental anomalies can be simultaneously satisfied, while being compatible with current constraints on the electron couplings of the $X$ boson.
This is the case of the KTeV anomaly in $\pi^0 \to e^+ e^-$ decay~\cite{KTeV:2006pwx}, inside the green region in Fig.~\ref{plot:vector}, and the
anomalous magnetic moment of the electron $(g-2)_e$, as we will explain in the following.

\subsubsection{KTeV anomaly and anomalous $(g-2)_e$}\label{sec:ktev}

\begin{figure}[t!]
\centering
{\includegraphics[width=.48\textwidth]{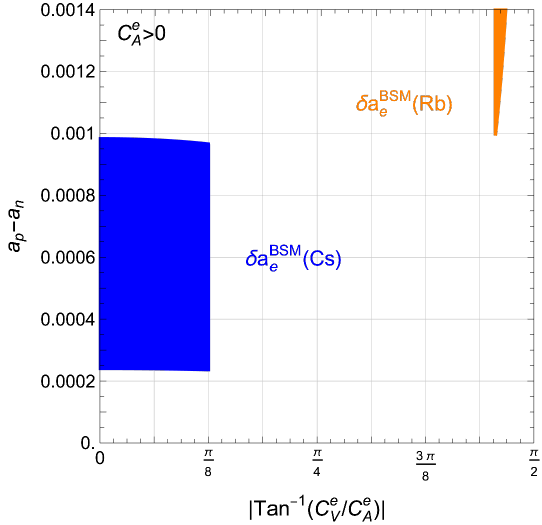}}\hfill
{\includegraphics[width=.48\textwidth]{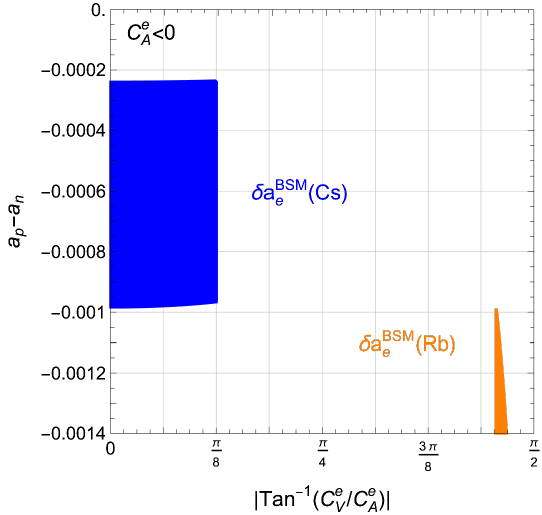}}  \\
\vspace{0.3cm}
\includegraphics[scale=0.75]{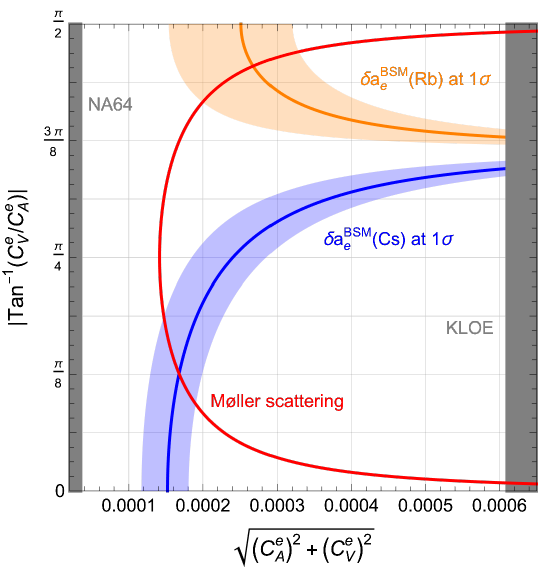}
\caption{{\it Upper panels:}
Values of the isovector nucleon axial coupling $a_p - a_n$ able to explain the KTeV anomaly at $1\sigma$ varying the ratio of vector and axial electron coupling for the two distinct cases $C_{A}^{e}>0$ and $C_{A}^{e}<0$.
{\it Lower panel:} Bounds on the vector and axial couplings of the electron to a spin-1 boson with mass $m_{X}\sim17$ MeV. The gray regions are excluded by NA64 and KLOE searches, while the region inside the red contour is excluded by M\o ller scattering. Here we assume ${\rm BR}(X\to e^+ e^-)=1$.
}
\label{plot:KTeV-g-2}
\end{figure}

 The lowest SM contribution to this decay is a one loop process with two photons as intermediate states. The KTeV-E779 Collaboration reports the measured value \cite{PhysRevD.75.012004}
\be
\text{BR}(\pi^{0}\to e^{+}e^{-})_{\text{exp}}=(7.48\pm0.29\pm0.25)\times10^{-8},
\ee
where they extrapolated from a selected  kinematic region to the entire one. The most recent calculation of SM prediction is \cite{PhysRevLett.128.172004}
\be
\text{BR}(\pi^{0}\to e^{+}e^{-})_{\text{SM}}=(6.25\pm0.03)\times10^{-8}.
\ee
The discrepancy of $3.2\sigma$\footnote{Using the latest radiative corrections from \cite{Vasko:2011pi,Husek:2014tna}, the full branching ratio extrapolated from the KTeV measurement is $\text{BR}(\pi^{0}\to e^{+}e^{-})_{\text{exp}}=(6.85\pm0.27\pm0.23)\times10^{-8}$, thus reducing the discrepancy to $1.8\sigma$.} could be explained assuming a BSM contribution from a light axial boson \cite{Kahn:2007ru}.
The actual best fit from the data reads as \cite{PhysRevLett.128.172004}
\be\label{eq:KTeVan}
\frac{(a_p-a_n)C^{e}_{A}}{g_A m_{X}^{2}}=2.60_{-1.60}^{+1.50}\times 10^{-10}\,\,\,\text{MeV}^{-2} \ ,
\ee
where $C_A^e$ is the axial coupling of the $X$ boson to the electron, see Eq.~\eqref{eq:lagV}.
A light vector contributes to the anomalous magnetic moment of the electron. The SM prediction from the measurement of the fine structure constant $\alpha$ from Cs atoms~\cite{Parker:2018vye} and the more recent prediction based on the measurement of $\alpha$ from Rb atoms~\cite{Morel:2020dww} are in contradiction among themselves. By asking that the BSM contribution from the $X$ boson 
given by~\cite{Jegerlehner:2009ry}
\be
\delta a_l^{\rm BSM} = \frac{C_V^2}{4\pi^2}\frac{m_\ell^2}{m_X^2} \frac{1}{2}
\int_0^1 {\rm d}z\; \frac{2 m_X^2 z^2 (z-1)}{m_X^2 (z-1) - m_\ell^2 z^2}+
\frac{C_A^2}{4\pi^2}\frac{m_\ell^2}{m_X^2} \frac{1}{2}
\int_0^1 {\rm d}z\; \frac{4 z^3 m_\ell^2  + 2 z m_X^2 (4-5z+z^2)}{m_X^2 (z-1) - m_\ell^2 z^2} \ ,
\ee
 doesn't overshoot the discrepancy between the central values of the SM prediction and the experimental measurement~\cite{Hanneke:2008tm} one obtains two different constraints, depending on the choice of the SM prediction
\begin{align}
& \delta a_e^{\rm BSM}(Rb) \simeq 7.6\times10^{-6} {C_V^e}^2 -3.80 \times 10^{-5 }  {C_A^e}^2  \in [ 0 - 0.48 \times 10^{-12} ]  \ , \\
& \delta a_e^{\rm BSM}(Cs) \simeq 7.6\times10^{-6} {C_V^e}^2 -3.80 \times 10^{-5 }  {C_A^e}^2  \in [  -0.88 \times 10^{-12} - 0 ]  \ .
\end{align}

The Cs atoms SM prediction naturally suggests a pure axial boson and the discrepancy observed in the electron anomalous magnetic moment
would be resolved at $1\sigma$ with an electron coupling
\be
C_{A}^{e}=\pm(1.52\pm0.31)\times10^{-4} \ .
\ee
By fixing this value for $C_A^e$, we have shown in Fig.~\ref{plot:vector} the parameter space of the nucleon couplings which can explain the KTeV anomaly see Eq.~\eqref{eq:KTeVan}, for the two distinct cases $C_{A}^{e}>0$ and $C_{A}^{e}<0$.

Allowing instead for both a vector and axial contribution to the electron coupling, in the upper panel of Fig~\ref{plot:KTeV-g-2} we show the values of the isovector nucleon axial coupling $a_p - a_n$ able to explain the KTeV anomaly at $1\sigma$ and assuming the discrepancy observed in the electron anomalous magnetic
 moment to be resolved for the Cs atoms (blue) and Rb atoms (orange) SM prediction for the two distinct cases $C_{A}^{e}>0$ and $C_{A}^{e}<0$. In the lower panel of the same figure 
 we show instead the most relevant bounds on this scenario with generic $X$ vector couplings to electrons, again for BR$(X\to e^+ e^-)=1$, which arise from the measurement of $e^+ e^-$ scattering from the KLOE experiment at DA$\Phi$Ne collider~\cite{Anastasi:2015qla}, measurements on parity violation in M\o ller scattering at SLAC~\cite{Anthony:2005pm} and beam dump experiment at NA64~\cite{Banerjee:2018vgk,Banerjee:2019hmi}, see App.~\ref{sec_pheno} for details. Interestingly, the PADME experiment will completely cover the region between the NA64 and the KLOE exclusions thus allowing for a strong test of the existence of the $X$ boson~\cite{Nardi:2018cxi,Darme:2022zfw}.

\subsubsection{Minimal SM extension with a new $U(1)$ gauge symmetry}

\begin{figure}[t!]
\centering
{\includegraphics[width=.49\textwidth]{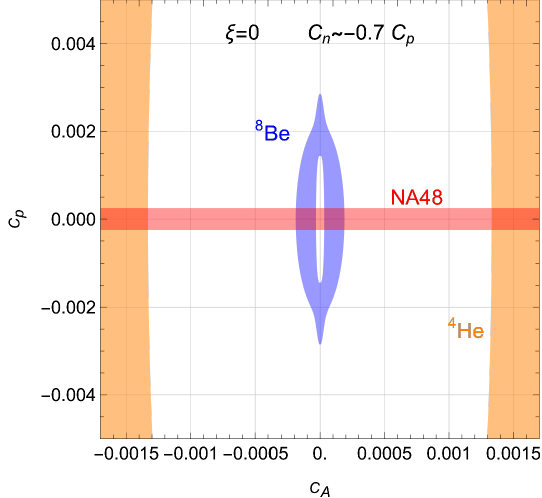}}\hfill
{\includegraphics[width=.49\textwidth]{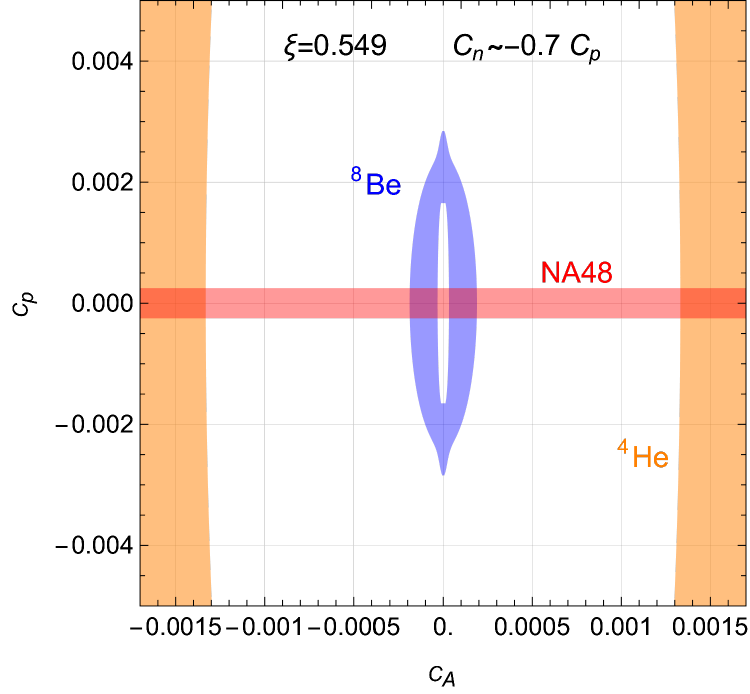}} 
\caption{
Regions of the $C_p-C_A$ couplings for the minimal SM extension with a new $U(1)_X$ symmetry where the $^{8}$Be (blue) and $^{4}$He (orange) anomalous ATOMKI transition can be explained at $1\sigma$. Inside the red region the NA48 bound is satisfied. In the left and right panel we assume $\xi=0$ and $\xi=0.549$ respectively, see App.~\ref{sec:em_ME} for details.}
\label{plot:mm}
\end{figure}

A common proposal in order to include a light vector $X$ in a SM extension is by enlarging its gauge group $G_{\rm SM}=SU(3)_{c}\times SU(2)_{W}\times U(1)_{Y}$ with a new abelian symmetry $U(1)_{X}$\footnote{The resulting model is typically not anomaly free, leading at low energy to stringent constraints for a light vector boson~\cite{Dror:2017ehi,Dror:2017nsg}. A possible way out has been described in~\cite{DiLuzio:2022ziu}. }.
In a minimal scenario, it is natural to assume that the SM Lagrangian is symmetric under $U(1)_{X}$. It follows that the $U(1)_{X}$ charges are a linear combination of the hypercharge $Y$ and all the accidental symmetries of the SM: the baryon number $B$ and the three lepton family numbers $L_{e,\mu,\tau}$.
When the gauge group $G_{\rm SM} \times U(1)_{X}$ is broken to $SU(3)_{c}\times U(1)_{Q}$, the $X$ field mixes with the other neutral gauge bosons in order to compose the physical states: the photon $\gamma$, the $Z^{0}$ boson and the light boson $X$. Moreover, once the symmetry is broken, the $X$ charges acquire a contribution from the diagonal weak isospin $T_{W_{3}}$.
In the quark sector, the $X$ couplings are then a linear combination of three independent charge assignments: $B$, $Y=Q-T_{W_{3}}$ and $T_{W_{3}}$. The baryon number $B$ and the electric charge $Q$ are vector symmetries so only the weak isospin $T_{W_{3}}$ induces an axial coupling. Thus, the axial couplings of the light quarks satisfy
\begin{equation}
C_{A}^{u}=-C_{A}^{d}=-C_{A}^{s}=-C_{A}^{e}\equiv C_{A}
\end{equation}
while the vector coupling for up and down quarks are independent. The nucleon couplings are obtained from the quark ones by Eq.~\eqref{nvc} and Eq.~\eqref{nac}.
However, the results from atomic parity violation experiments~\cite{Porsev:2009pr}, see App.~\ref{sec_pheno}, strongly constrain these couplings, and require for them (product of) values so small that the ATOMKI anomalies cannot be explained, see Eq.~\eqref{apv}.
A way to avoid this bound is to assume somehow a magical cancellation between the up and down vector couplings, which is
\be
C_{V}^{d}=-\frac{188}{211}C_{V}^{u} \ ,
\ee
thus
\be
C_{n}=-\frac{55}{78}C_{p}\simeq-0.7\,C_{p} \ .
\ee
Hence, we are left with two independent couplings, $C_{p}$ and $C_{A}$. 
However, as we show in Fig.~\ref{plot:mm}, there is no possible simultaneously explanation of the $^8$Be and $^4$He anomalies in the minimal BSM scenario considered here for both choices of the isospin breaking parameter $\xi$, which produce almost  indistinguishable results.

%%%%%%%%%%%%%%%%%%%%%%%%%%%%%%%%%%%%%%%%%%%%%%%%%%%%%%%%%%%%%

\section{Conclusions}\label{sec:conc}

Motivated by the latest experimental results recently released by the ATOMKI collaboration, we have critically re-examined the possible theoretical interpretation of the observed anomalies in $^8$Be, $^4$He and $^{12}$C anomalies in terms of a BSM boson $X$ with mass $\sim 17\;$MeV. After having reviewed the current status of the ATOMKI results and the kinematic of the observed excesses we have employed a multipole expansion formalism to compute the anomalous decay rate for the decay of the excited nuclei into an $e^+ e^-$ pair via an intermediate on-shell BSM state.
Our results identify an axial vector state as the most promising candidate to simultaneously explain all the three anomalous nuclear decay, while the other spin/parity assignments seems disfavored for a combined explanation.
However, the axial nuclear matrix element of the $^{12}$C transition is currently unknown and our conclusions are based on an order of magnitude estimate for its value. Before being able to make a definite claim regarding the possibility of combined explanation of the ATOMKI anomalies with an axial vector state, this matrix elements need to be evaluated.
Intriguingly, the hypothesis of an axial vector state can also simultaneously accommodate other experimental anomalies, as the one observed by the KTeV experiment in $\pi^0\to e^+ e^-$ decay, while being compatible with the conflicting measurements of the anomalous magnetic moment of the electron $(g-2)_e$ and other experimental constraints.
The independent experimental tests that will be performed by the MEG II experiment~\cite{MEGII:2018kmf} at PSI and the by the Montreal Tandem accelerator~\cite{Azuelos:2022nbu} will definitely 
answer the question regarding the nature of the ATOMKI results and tell us whether these anomalies are merely due to unaccounted SM and/or experimental effects or else are the first signs of the long sought new physics beyond the Standard Model.

%%%%%%%%%%%%%%%%%%%%%%%%%%%%%%%%%%%%%%%%%%%%%%%%%%%%%%%%%%%%%
%%%%%%%%%%%%%%%%%%%%%%%%%%%%%%%%%%%%%%%%%%%%%%%%%%%%%%%%%%%%%

\section*{Acknowledgments}

The Authors thank Marco Nardecchia and Mauro Raggi for triggering and encouraging this work. Our gratitude to Luca Di Luzio, Gabriel Massoni Salla and Federico Ripani for useful comments and discussions.
We are also grateful to Matheus Hostert and Maxim Pospelov who helped us to spot a mistake in the propagation of the errors for the best fit of the vector state which was present in the published version of our paper.
The work of CT was supported in part by MIUR under contract PRIN 2017L5W2PT.

%%%%%%%%%%%%%%%%%%%%%%%%%%%%%%%%%%%%%%%%%%%%%%%%%%%%%%%%%%%%% APPENDICES
%%%%%%%%%%%%%%%%%%%%%%%%%%%%%%%%%%%%%%%%%%%%%%%%%%%%%%%%%%%%%

\appendix

%%%%%%%%%%%%%
%%%% TABLES %%%%
%%%%%%%%%%%%

\section{Additional tables and input parameters}\label{sec:app_num_table}

Here we report the numerical values for the nuclear transition of interest used throughout our analysis.

\begin{table}[h!]
\begin{center}
\scalebox{0.8}{
\begin{tabular}{ccccccc}
\toprule
$E_{b}$\;[MeV] & $A$ & $m_{A}$\;[MeV] & $m_{pA}/m_{p}$ & $N^{*}$ & $m_{N^{*}}$\;[MeV] & $v_{N^{*}}\,/\,c$ \\
\midrule
1.03 & $^{7}$Li & 6533.83 & 0.87 & $^{8}$Be(18.15) & 7473.01 & 0.0059 \\
\midrule
0.45 & $^{7}$Li & 6533.83 & 0.87 & $^{8}$Be(17.64) & 7472.50 & 0.0039 \\
\midrule
1.59 & $^{3}$H & 2808.92 & 0.75 & $^{4}$He(21.01) & 3748.39 & 0.0146 \\
\midrule
0.52 & $^{3}$H & 2808.92 & 0.75 & $^{4}$He(20.21) & 3747.59 & 0.0084 \\
\midrule
1.40 & $^{11}$B & 10252.54 & 0.92 & $^{12}$C(17.23) & 11192.09 & 0.0046 \\
\bottomrule
\end{tabular}
}
\end{center}
\caption{Proton kinetic energy $E_b$, mass of the target and of the excited nucleus, reduced proton-target mass and velocity of the excited nucleus produced by the collision in the lab frame at the resonance peak ($E_{CM}=m_{N^{*}}$) for all the measured $N^{*}$ states.}
\label{tab:kin:1}
\end{table}

\begin{table}[h!]
\begin{center}
\scalebox{0.8}{
\begin{tabular}{ccccc}
\toprule
$N^{*}$ & $m_{N}$\;[MeV] & $E_{th}$\;[MeV] & $\omega$\;[MeV] & $v_{X}$ \\
\midrule
$^{8}$Be(18.15) & 7454.86 & 17.24 & 18.15 & 0.350 \\
\midrule
$^{8}$Be(17.64) & 7454.86 & 17.24 & 17.64 & 0.267 \\
\midrule
$^{4}$He(21.01) & 3727.38 & 19.81 & 21.01 & 0.588 \\
\midrule
$^{4}$He(20.21) & 3727.38 & 19.81 & 20.21 & 0.541 \\
\midrule
$^{12}$C(17.23) & 11174.86 & 15.95 & 17.23 & 0.163 \\
\bottomrule
\end{tabular}
}
\end{center}
\caption{$N$ mass, threshold energy $E_{th}$, boson energy $\omega$ in the CM frame the velocity $v_{X}$ of an hypothetical boson of mass $m_{X}\simeq 17$ MeV emitted in the CM frame at the resonance peak ($E_{CM}=m_{N^{*}}$) for all the measured $N^{*}$ states.}
\label{tab:kin:2}
\end{table}

\begin{table}[h!]
\begin{center}
\scalebox{0.8}{
\begin{tabular}{cccc}
\toprule
$N^{*}$ & $\delta$ & $y_{\rm max}$ & $\theta_{\pm}^{\rm min}$[$^\circ$]\\
\midrule
$^{8}$Be(18.15) & 0.056 & 0.351 & 139.0 \\
\midrule
$^{8}$Be(17.64) & 0.058 & 0.267 & 149.0 \\
\midrule
$^{4}$He(21.01) & 0.049 & 0.589 & 107.9 \\
\midrule
$^{4}$He(20.49) & 0.050 & 0.559 & 112.1 \\
\midrule
$^{4}$He(20.21) & 0.051 & 0.542 & 114.4 \\
\midrule
$^{12}$C(17.23) & 0.059 & 0.163 & 161.2 \\
\bottomrule
\end{tabular}
}
\end{center}
\caption{$\delta$ parameter, maximum energy asymmetry $y_{max}$ and minimal opening angle $\theta_{\pm}^{min}$ for an hypothetical boson of mass $m_{X}=17$ MeV emitted in the CM frame.}
\label{tab:kin:3}
\end{table}

%%%%%%%%%%%%%%%%%%%%%
%%% EM processes %%%%%%%%%%
%%%%%%%%%%%%%%%%%%%%%

\newpage

\section{Electromagnetic dynamics}
\label{app:sec_EM}

With the formalism described in Sec.~\ref{sec:signal} we can describe the dynamics for the SM processes pictured in Fig.~\ref{fig:diagrams} of  real $\gamma$ emission and IPC.
The nuclear electromagnetic current, including the anomalous magnetic moments, is given by
\be
\mathcal{J}_{\mu}^{(\gamma)}=eQ_{p}\bar{p}\gamma_{\mu}p+eQ_{n}\bar{n}\gamma_{\mu}n+\frac{e\kappa_{p}^{\gamma}}{2m_{p}}\partial^{\nu}(\bar{p}\sigma_{\mu\nu}p)+\frac{e\kappa_{n}^{\gamma}}{2m_{n}}\partial^{\nu}(\bar{n}\sigma_{\mu\nu}n),
\ee
where $\kappa_p = +1.792847351(28)$, $\kappa_n = -1.9130427(5)$~\cite{book:80102} and $Q_{p,n}$ indicates the electric charge of the nucleon in units of the absolute electron charge. 
The magnetic momenta of the nucleons are
\be
\mu_{p,n}=(Q_{p,n}+\kappa_{p,n}^{\gamma})\mu_{\N} \ ,
\ee
where $\mu_{\N}=e/2m_{\N}$ is the nuclear magneton.
The conservation of the electromagnetic current implies that only three independent spherical operators have to be considered, {\emph{c.f.r.}} Eq.~\eqref{LtoM}. In the non relativistic and long wavelength approximation the spherical operators with $J=0,1$ are given by
\begin{align}
\mathcal{M}_{00}^{(\gamma)} & \simeq -\frac{ek^{2}}{6}\rho^{(\gamma)} \ , \\
\mathcal{M}_{1M}^{(\gamma)} & \simeq \frac{ek}{3}d_{M}^{(\gamma)} \ , \\
\mathcal{T}_{1M}^{\rm el(\gamma)} & \simeq \frac{\sqrt{2}e\omega}{3}d_{M}^{(\gamma)} \ , \\
\mathcal{T}_{1M}^{\rm mag(\gamma)} & \simeq \frac{i\sqrt{2}k\mu_{\N}}{3}\mu_{M}^{(\gamma)} \ , 
\end{align}
where we have defined the electromagnetic monopole $\rho^{(\gamma)}$, the electric dipole $d_M^{(\gamma)}$ and the magnetic moment $\mu_M^{(\gamma)}$ operators as 
\begin{align}
\rho^{(\gamma)} &= \frac{1}{\sqrt{4\pi}}\sum_{s=1}^{A}Q_{s}r_{s}^{2} \ , \\
d_{M}^{(\gamma)} &= \sqrt{\frac{3}{4\pi}}\sum_{s=1}^{A}Q_{s}\vec{r}_{s}\cdot \hat{e}_{M} \ , \\
\mu_{M}^{(\gamma)} &= \sqrt{\frac{3}{4\pi}}\sum_{s=1}^{A}[Q_{s}(\vec{r}_{s}\times\vec{p}_{s})+(Q_{s}+\kappa_{s}^{\gamma})\vec{\sigma}_{s}]\cdot \hat{e}_{M} \ .
\end{align}

\subsection*{Real $\gamma$ emission}

The rate for the process with a real $\gamma$ emission can be readily computed from Eq.~\eqref{dlcc} by fixing $m_\gamma=0$. In this case, due to the transversality of the photon, only processes with $J=0$ are allowed which are
\begin{itemize}
\item electric type transitions $EJ$ from the contribution of $\mathcal{T}_{J}^{\rm el}$ with parity $\pi(EJ)=(-1)^{J}$ , 
\item magnetic type transitions  $MJ$ from the contribution of $\mathcal{T}_{J}^{\rm mag}$ with parity $\pi(MJ)=(-1)^{J+1}$ . 
\end{itemize}
The $E1$ and $M1$ decay rates are equal to
\begin{gather}
\Gamma_{\gamma}^{E1}=\frac{16\pi\alpha\omega^{3}}{9(2J_{*}+1)}|\braket{f||d^{(\gamma)}||i_{*}}|^{2}, \label{geE1}\\
\Gamma_{\gamma}^{M1}=\frac{4\mu_{\N}^{2}\omega^{3}}{9(2J_{*}+1)}|\braket{f||\mu^{(\gamma)}||i_{*}}|^{2}.
\end{gather}

\subsection*{Internal pair creation}

\begin{figure}[t!]
\begin{center}
\includegraphics[scale=0.95]{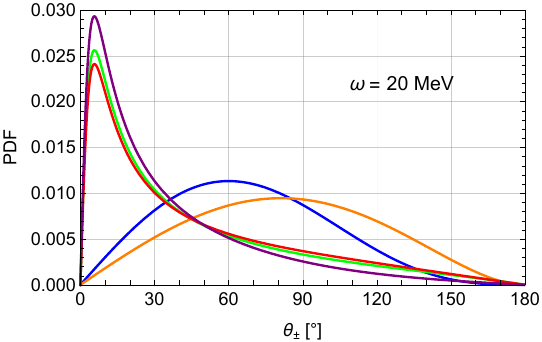}
\caption{Normalized angular correlation distributions of the $e^+ e^-$ pair from the various multipole contribution to the IPC process: $L0$ (blue), $L1$ (orange), $E1$ (green), $E1+L1$ (red), $M1$ (purple).}
\label{plot:theta-em}
\end{center}
\end{figure}

At lowest order the IPC process involves the emission of a virtual photon decaying into an $e^+ e^-$ pair. The differential decay rate with respect to the energy asymmetry $y$ and the opening angle $\theta_\pm$ is given by
\be
\begin{split}
\frac{d^{2}\Gamma_{\pm}}{dy\,d\cos\theta_\pm}&=\frac{2\omega}{2J_{*}+1} \frac{\alpha}{4\pi} \Biggl\{ f_{\mathcal{M}}(y,\cos\theta_\pm,\delta)\sum_{J\geq 0}\left|\braket{f||\mathcal{M}_{J}||i_{*}}\right|^{2}\\
&+ f_{\mathcal{T}}(y,\cos\theta_\pm,\delta)\sum_{J\geq 1}\left[\left|\braket{f|| \mathcal{T}_{J}^{el}||i_{*}}\right|^{2} + \left|\braket{f||\mathcal{T}_{J}^{mag}||i_{*}}\right|^{2}\right]\Biggr\} \ ,
\end{split}
\ee
where the $f_{\mathcal{M}, \mathcal{T}}(y,\cos\theta_\pm,\delta)$ functions are 
\begin{gather}
f_{\mathcal{M}}(y,c,\delta)=\frac{\sqrt{(1-\delta^{2}+y^{2})^{2}-4y^{2}}\left[1-y^{2}-\delta^{2}+c\sqrt{(1-\delta^{2}+y^{2})^{2}-4y^{2}}\right]}{\left[1+y^{2}-\delta^{2}+c\sqrt{(1-\delta^{2}+y^{2})^{2}-4y^{2}}\right]^{2}},\\
\begin{split}
&f_{\mathcal{T}}(y,c,\delta)=\frac{1}{2}\frac{\sqrt{(1-\delta^{2}+y^{2})^{2}-4y^{2}}}{\left[1+y^{2}-\delta^{2}+c\sqrt{(1-\delta^{2}+y^{2})^{2}-4y^{2}}\right]}\times\\
\times&\frac{\left[\left(1-3y^{2}+3\delta^{2}-c\sqrt{(1-\delta^{2}+y^{2})^{2}-4y^{2}}\right)\left(1+y^{2}-\delta^{2}+c\sqrt{(1-\delta^{2}+y^{2})^{2}-4y^{2}}\right)+4y^{2}\right]}{\left[1-y^{2}+\delta^{2}-c\sqrt{(1-\delta^{2}+y^{2})^{2}-4y^{2}}\right]^{2}}.
\end{split}
\end{gather}
Note that IPC processes acquire a contribution also from the longitudinal modes, absent in the real $\gamma$ emission case. Hence, processes of pair production are of three types:
\begin{itemize}
\item longitudinal type transition $LJ$ from the contribution of $\mathcal{M}_{J}$ with parity $\pi(LJ)=(-1)^{J}$,
\item electric transition $EJ$ from the contribution of $\mathcal{T}_{J}^{\rm el}$ with parity $\pi(EJ)=(-1)^{J}$,
\item magnetic transition $MJ$ from the contribution of $\mathcal{T}_{J}^{\rm mag}$ with parity $\pi(MJ)=(-1)^{J+1}$ \ ,
\end{itemize}
with differential decay rates\footnote{$\zeta(y, \delta, \cos\theta_\pm) = 1+y^{2}-\delta^{2}+\cos\theta_\pm \sqrt{(1-\delta^{2}+y^{2})^{2}-4y^{2}}$}
\begin{align}
\frac{d^{2}\Gamma_{\pm}^{L0}}{dy\,d\cos\theta_\pm} & = \frac{\alpha^{2}\omega^{5}}{72(2J_{*}+1)}
\zeta^2(y,\delta,\cos\theta_\pm)f_{\mathcal{M}}(y,\cos\theta_\pm,\delta)  |\braket{f||\rho^{(\gamma)}||i_{*}}|^{2} \ , \label{ppL0} \\
\frac{d^{2}\Gamma_{\pm}^{L1}}{dy\,d\cos\theta_\pm} & = \frac{\alpha}{16\pi} \zeta(y,\delta,\cos\theta_\pm)f_{\mathcal{M}}(y,\cos\theta_\pm,\delta) \Gamma_{\gamma}^{E1} \ , \\
\frac{d^{2}\Gamma_{\pm}^{E1}}{dy\,d\cos\theta_\pm}& = \frac{\alpha}{4\pi}  f_{\mathcal{T}}(y,\cos\theta_\pm,\delta)\Gamma_{\gamma}^{E1} \ , \\
\frac{d^{2}\Gamma_{\pm}^{M1}}{dy\,d\cos\theta_\pm} &= \frac{\alpha}{8\pi} \zeta(y,\delta,\cos\theta_\pm) f_{\mathcal{T}}(y,\cos\theta_\pm,\delta) \Gamma_{\gamma}^{M1} \ .
\end{align}
We show in Fig.~\ref{plot:theta-em} the theoretical distributions of the angular correlation, obtained after integrating the above expressions over the asymmetry $y$.

\begin{figure}[t!]
\begin{center}
\includegraphics[scale=0.95]{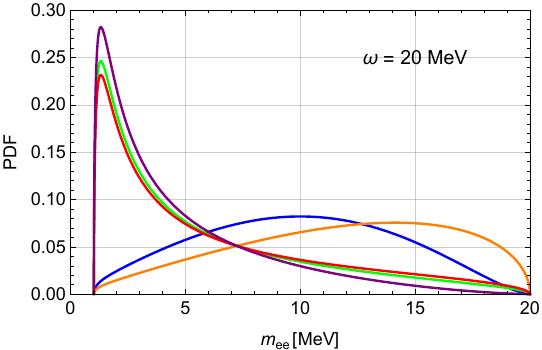}
\caption{Normalized invariant mass distributions of the $e^+ e^-$ pair from the various multipole contribution to the IPC process: $L0$ (blue), $L1$ (orange), $E1$ (green), $E1+L1$ (red), $M1$ (purple).}
\label{plot:mee:1}
\end{center}
\end{figure}

The reduced invariant mass of the lepton pair is a function of the energy asymmetry and opening angle. Given the (normalized) distribution function $f(y,\cos\theta_\pm,\delta)$ of the pairs over the plane defined by $y$ and $\cos\theta_\pm$, the (normalized) distribution function $g(s,\delta)$ of the reduced invariant mass is given by
\be
\begin{split}
g(s,\delta)&=\int_{-1+\delta}^{1-\delta}\!dy\,\int_{-1}^{1}\!d\cos\theta_\pm\,\delta\!\left(s-1+y^{2}-\delta^{2}+\cos\theta_\pm\sqrt{(1-\delta^{2}+y^{2})^{2}-4y^{2}}\right)\,f(y,\cos\theta_\pm,\delta)\\
&=\int_{-\sqrt{(2-s)(s-2\delta^{2})/2s}}^{\sqrt{(2-s)(s-2\delta^{2})/2s}}\!dy\,\frac{1}{\sqrt{(1-\delta^{2}+y^{2})^{2}-4y^{2}}}f\left(y,\frac{1-y^{2}+\delta^{2}-s}{\sqrt{(1-\delta^{2}+y^{2})^{2}-4y^{2}}},\delta\right) \ ,
\end{split}
\ee
where $s=\frac{2m_{ee}^2}{\omega^2}\;(2\delta^2 \le s \le 2)$ is the reduced invariant mass.
For the $J=0,1$ multipoles we calculated, the integration over the energy asymmetry is easily performed since the integrand turns out to be polynomial in $y$. Hence, one finds
\begin{gather}
g^{L0}(s,\delta)=\mathcal{N}_{L0}(\delta)\,(2-s)^{3/2}s^{-3/2}(s-2\delta^{2})^{1/2}(s+\delta^{2})\nonumber\\
\text{with}\,\,\,\,\,\,\,\,\mathcal{N}_{L0}^{-1}(\delta)=\int_{2\delta^{2}}^{2}\!dx\,(2-x)^{3/2}x^{-3/2}(x-2\delta^{2})^{1/2}(x+\delta^{2})\ ,\\
g^{L1}(s,\delta)=\mathcal{N}_{L1}(\delta)\,(2-s)^{1/2}s^{-3/2}(s-2\delta^{2})^{1/2}(s+\delta^{2})\nonumber\\
\text{with}\,\,\,\,\,\,\,\,\mathcal{N}_{L1}^{-1}(\delta)=\int_{2\delta^{2}}^{2}\!dx\,(2-x)^{1/2}x^{-3/2}(x-2\delta^{2})^{1/2}(x+\delta^{2})\ ,\\
g^{E1}(s,\delta)=\mathcal{N}_{E1}(\delta)\,(2-s)^{1/2}s^{-5/2}(s-2\delta^{2})^{1/2}(s+\delta^{2})\nonumber\\
\text{with}\,\,\,\,\,\,\,\,\mathcal{N}_{E1}^{-1}(\delta)=\int_{2\delta^{2}}^{2}\!dx\,(2-x)^{1/2}x^{-5/2}(x-2\delta^{2})^{1/2}(x+\delta^{2})\ ,\\
g^{E1+L1}(s,\delta)=\mathcal{N}_{E1+L1}(\delta)\,(2-s)^{1/2}s^{-5/2}(s-2\delta^{2})^{1/2}(s+\delta^{2})(s+4)\nonumber\\
\text{with}\,\,\,\,\,\,\,\,\mathcal{N}_{E1+L1}^{-1}(\delta)=\int_{2\delta^{2}}^{2}\!dx\,(2-x)^{1/2}x^{-5/2}(x-2\delta^{2})^{1/2}(x+\delta^{2})(x+4)\ ,\\
g^{M1}(s,\delta)=\mathcal{N}_{M1}(\delta)\,(2-s)^{3/2}s^{-5/2}(s-2\delta^{2})^{1/2}(s+\delta^{2})\nonumber\\
\text{with}\,\,\,\,\,\,\,\,\mathcal{N}_{M1}^{-1}(\delta)=\int_{2\delta^{2}}^{2}\!dx\,(2-x)^{3/2}x^{-5/2}(x-2\delta^{2})^{1/2}(x+\delta^{2})\ .
\end{gather}
We show in Fig.~\ref{plot:mee:1} their theoretical distributions.

\subsection*{Isospin mixing and electromagnetic nuclear matrix elements}
\label{sec:em_ME}

The $^{8}\text{Be}(18.15)$ and $^{8}\text{Be}(17.64)$ states, close in energy and with same spin-parity assignment, presents a significant isospin mixing. In general, given a doublet of nuclear energy levels of spin $J$ with mixed isospin, the physical states (denoted with $a$ and $b$) are given by a linear combination of states with fixed isospin
\begin{equation}
\Psi_{a}^{J}=\alpha_{J}\Psi_{I=0}^{J}+\beta_{J}\Psi_{I=1}^{J},\,\,\,\,\,\Psi_{b}^{J}=-\alpha_{J}\Psi_{I=1}^{J}+\beta_{J}\Psi_{I=0}^{J}
\end{equation}
where $a$ labels the lowest energy level between them. The coefficients $\alpha_{J}$ and $\beta_{J}$ are real and satisfy $\alpha_{J}^{2}+\beta_{J}^{2}=1$. For the $^{8}$Be nucleus, the values of the mixing coefficients have been evaluated through Quantum Monte Carlo simulation \cite{Pastore:2014oda}. The result for the $J=1$ doublet is
\begin{equation}
\alpha_{1}=0.21(3),\,\,\,\,\beta_{1}=0.98(1).
\end{equation}
According to this, we define the isospin magnetic strength $M1_{I=0,1}^{\gamma}$ by
\begin{equation}
\begin{split}
\braket{^{8}\text{Be}||\mu^{(\gamma)}||^{8}\text{Be}(17.64)}&=\alpha_{1}M1_{I=0}^{\gamma}+\beta_{1}M1_{I=1}^{\gamma},\\
\braket{^{8}\text{Be}||\mu^{(\gamma)}||^{8}\text{Be}(18.15)}&=-\alpha_{1}M1_{I=1}^{\gamma}+\beta_{1}M1_{I=0}^{\gamma}.
\end{split}
\end{equation}
whose values has been estimated to be \cite{Pastore:2014oda}
\begin{equation}
M1_{I=0}^{\gamma}=0.014(1),\,\,\,\,\,M1_{I=1}^{\gamma}=0.767(9).
\end{equation}
At this level, a direct comparison with the experimental values of the decay lengths shows a significant discrepancies with the theoretical prediction. Following \cite{Feng:2016ysn}, we'll consider the deficiency as due to isospin breaking effects we neglected in the first attempt. The inclusion of them is obtained trough the introduction of a $\Delta I=1$ spurion, whose effective result is to shift the nuclear matrix elements as
\be
\begin{split}
\braket{^{8}\text{Be}||\mu^{(\gamma)}||^{8}\text{Be}(17.64)}&=\alpha_{1}M1_{I=0}^{\gamma}+\beta_{1}M1_{I=1}^{\gamma}+\alpha_{1}\xi M1_{I=1}^{\gamma},\\
\braket{^{8}\text{Be}||\mu^{(\gamma)}||^{8}\text{Be}(18.15)}&=-\alpha_{1}M1_{I=1}^{\gamma}+\beta_{1}M1_{I=0}^{\gamma}+\beta_{1}\xi M1_{I=1}^{\gamma}.
\end{split}
\ee
The parameter $\xi$ characterizes the strength of the spurion and its size is controlled by non-perturbative effects. One finds $\xi=0.549$ by requiring that the resulting decay width $\Gamma(^{8}\text{Be}(17.64)\to \,^{8}\text{Be} +\gamma)$ reproduces its experimental value. For $\xi=0$ the isospin breaking effects are simply neglected.
\\
Parity conservation prohibits electromagnetic interaction in the $^{4}\text{He}(21.01)$ transition to the ground state, thus the nuclear matrix element relative to this decay is equal to zero.\\
Due to the massless nature of the photon, the $\gamma$-emission decay width of the $^{4}\text{He}(20.21)$ transition to the ground state also vanishes but the IPC process is still possible. Pair production is mediated by the monopole operator $\rho^{(\gamma)}$, whose matrix element has been measured to be \cite{Walcher:1970vkv}
\be
\sqrt{4\pi}\braket{ ^{4}\text{He}||\rho^{(\gamma)}|| ^{4}\text{He}(20.21)}=(1.10\pm0.16)\,\text{fm}^{2},
\ee
with corresponding decay length, after integrating \eqref{ppL0}, equal to $\Gamma_{\pm}=(3.3\pm1.0)\times10^{-4}$ eV.\\
\\
The $\gamma$-emission decay length of the $^{12}$C(17.23) transition to the ground state has been measured to be $\Gamma_{\gamma}=44$ eV \cite{Segel:1965zz}. Hence, from \eqref{geE1}, one finds
\be
\braket{ ^{12}\text{C}||d^{(\gamma)}|| ^{12}\text{C}(17.23)}=0.157\,\text{fm} \ .
\ee

%%%%%%%%%%%%%%%%%%%%%%%%%%%%%%%%%%%%%%%%%%%%%%%%%%%%%%%%%%%%%

%%%%%%%%%%%%%%%%%%%%%
%%%%%
%%%%%%%%%%%%%%%%%%%%%%

\section{Nucleon effective couplings}\label{qton}

We report in this section the matching between the effective interaction of the $X$ boson with the nuclear matter and its interactions with the fundamental SM degrees of freedom, quark and gluons. Since nucleons are spin 1/2 particles, for an operator ${\cal O}$ composed by quark fields one has
\begin{equation}
\label{matel}
\langle N, p'|\mathcal{O}|N, p\rangle=\bar{u}_{p'}\Gamma(p',p)u_{p} \ ,
\end{equation}
where $\Gamma$ is a matrix with spinor indices and $u_{p}$ is the solution of the free Dirac equation\footnote{This leaves the matrix element unchanged upon the substitution $\Gamma(p',p)\rightarrow\frac{\slashed{p}'+m}{2m}\Gamma(p',p)\frac{\slashed{p}+m}{2m}$.}. Lorentz invariance as well $C$, $P$ and $CPT$ symmetries impose further constraints on this matrix element. As mentioned in the main text, since the transferred momentum in the considered processes is generally much smaller than $\Lambda_{\rm QCD}$ we approximated the form factors that are in general present in these expressions as constants.

\subsection*{Vector interaction}

In the UV an $X$ vector boson interacts with a quark current of the form $C_{V}^{q}\bar{q}\gamma^{\mu}q$, which gives an effective interaction
\begin{equation}
{\cal L} = C_{N}\bar{N}\gamma^{\mu}N X_\mu+\frac{\kappa_{N}}{2m_{N}}\partial_{\nu}(\bar{N}\sigma^{\mu\nu}N) X_\mu+\frac{g_{N}}{m_{N}}\partial^{\mu}(\bar{N}N) X_\mu \ .
\end{equation}
Conservation of the vector current implies $g_{N}=0$,
while symmetry considerations fix
\begin{equation}
\label{nvc}
C_{p}=2C_{V}^{u}+C_{V}^{d} \ , \quad C_{n}=C_{V}^{u}+2C_{V}^{d} \ .
\end{equation}
Symmetry considerations don't allow to simplify the expression for the magnetic moments of the nucleons
\begin{equation}
\mu_{N}^{(X)}=\frac{(C_{N}+\kappa_{N})}{e}\,\mu_{\mathcal{N}} \ ,
\end{equation}
where $\mu_{\cal N}$ is the Bohr magneton, 
since even the sea quarks can give 
a contribution to these quantities. However by using the static quark model one can make an estimation
~\cite{dispense:2018}. Working under the assumption that the valence quarks of the nucleons have mass equal to $m_{u}\simeq m_{d}\simeq m_{\mathcal{N}}/3$\footnote{Here we are considering the effective mass of the quarks when they are bounded together by gluons, not their real mass.}, at lowest order the magnetic moment of each quark is given only by its charge and effective mass, {\it i.e.}
\be
\mu_{q}=\frac{C_{V}^{q}}{2m_{q}} \ .
\ee
In the static quark model one thus finds
\begin{align}
\label{qmmm}
& \mu_{p}^{(X)}=\langle p|\mu|p\rangle=\frac{4}{3}\mu_{u}^{(X)}-\frac{1}{3}\mu_{d}^{(X)}\simeq\frac{4C_{V}^{u}-C_{V}^{d}}{e}\,\mu_{\mathcal{N}}=\frac{3C_{p}-2C_{n}}{e}\,\mu_{\mathcal{N}} \ ,\\
& \mu_{n}^{(X)}=\langle n|\mu|n\rangle=-\frac{1}{3}\mu_{u}^{(X)}+\frac{4}{3}\mu_{d}^{(X)}\simeq\frac{-C_{V}^{u}+4C_{V}^{d}}{e}\,\mu_{\mathcal{N}}=\frac{-2C_{p}+3C_{n}}{e}\,\mu_{\mathcal{N}} \ .
\end{align}
For the electromagnetic couplings, $C_{p}=+e$ and $C_{n}=0$, one obtains values close to the experimental ones, $\mu_{p}^{{\rm em}}|_{\rm exp}\simeq +2.792\,\mu_{\mathcal{N}}$ and $\mu_{n}^{{\rm em}}|_{{\rm exp}}\simeq-1.913\,\mu_{\mathcal{N}}$, within a $10\%$. The interaction between quarks and the $X$ particle might also come from an effective magnetic moment interaction ${\cal L}= \frac{\kappa_q}{\Lambda} \partial_{\nu}(\bar{q}\sigma^{\mu\nu}q) X_\mu$ generated, {\it e.g.}, at loop level by integrating out some heavy particle in the low energy limit.
This effective operator contributes to the magnetic moment of the nucleons through the substitution
\begin{equation}
\partial_{\nu}(\bar{q}\sigma^{\mu\nu}q)\rightarrow\delta_{q}^{(N)}\partial_{\nu}(\bar{N}\sigma^{\mu\nu}N)
\end{equation}  
where $N=p,n$, which shifts $\kappa_N\to \kappa_N +  \frac{2 m_N}{\Lambda}\delta_q^{(N)}$ for each $q$. By lattice computation \cite{Dolgov:2002zm,Aoki:1996pi}, it has been estimated
\begin{equation}
\delta_{u}^{(p)}=\delta_{d}^{(n)}=0.84 \ , \quad \delta_{d}^{(p)}=\delta_{u}^{(n)}=-0.23\ , \quad\delta_{s}^{(p)}=\delta_{s}^{(n)}=-0.046
\end{equation}
for the light quarks contributions.

\subsection*{Axial interaction}

In the UV an $X$ axial vector boson interacts with a quark current of the form $C_{V}^{q}\bar{q}\gamma^{\mu}\gamma^5 q$
which brings to an effective nucleon current
\begin{equation}
{\cal L} = a_{N}\bar{N}\gamma^{\mu}\gamma^{5}N X_\mu+\frac{b_{N}}{m_{N}}\partial^{\mu}(i\bar{N}\gamma^{5}N) X_\mu+\frac{d_{N}}{2m_{N}}\partial_{\nu}(i\bar{N}\sigma^{\mu\nu}\gamma^{5}N) X_\mu\ .
\end{equation}
CP conservation in QCD interactions forces $d_N=0$, while the term proportional to $b_N$ doesn't contribute to the considered processes when one has on-shell $X$\footnote{This can be seen by performing and integration by parts.}. The nucleon axial couplings $a_{N}$ are given by the sum of quark coupling $a_{q}$ weight by the fraction of the spin of the nucleon $\Delta_{q}^{(N)}$,
\begin{equation}
\label{nac}
a_{N}=\sum_{q}\Delta_{q}^{(N)}C_{A}^{q} \ .
\end{equation}
These fractions are given by integrals of helicity-dependent parton distributions and can be measured in lepton nucleon scattering. Their values are equal to \cite{Kozaczuk:2016nma,Bishara:2016hek}
\begin{equation}
\Delta_{u}^{(p)}=\Delta_{d}^{(n)}=0.897(27) \ , \quad \Delta_{d}^{(p)}=\Delta_{u}^{(n)}=-0.367(27) \ , \quad\Delta_{s}^{(p)}=\Delta_{s}^{(n)}=-0.026(4) \ ,
\end{equation}
while the contributions from heavy quark are small and can be neglected. As for the vector case, it's possible that the interaction between quarks and the $X$ boson comes from an effective interaction like ${\cal L} = \frac{d_q}{\Lambda}\partial_{\nu}(i\bar{q}\sigma^{\mu\nu}\gamma^{5}q) X_\mu$.
This effective operator generates an electric dipole for the nucleons through the substitution
\begin{equation}
\partial_{\nu}(i\bar{q}\sigma^{\mu\nu}\gamma^{5}q)\rightarrow\delta_{q5}^{(N)}\partial_{\nu}(i\bar{N}\sigma^{\mu\nu}\gamma^{5}N)
\end{equation}  
where $N=p,n$, which again shifts the $d_N$ value as before. Unfortunately the values of $\delta_{q5}^{(N)}$ are difficult to be measured and are poor known. Only recently \cite{Anselmino:2008jk} it has been measured the light quark contribution to the proton at $Q^{2}=0.8$ GeV$^{2}$
\begin{equation}
\delta_{u5}^{(p)}=0.54_{-0.22}^{+0.09} \ , \quad \delta_{d5}^{(p)}=-0.23_{-0.16}^{+0.09} \ .
\end{equation}

\subsection*{Scalar interaction}

The scalar interaction between quarks and a spin $0$ particle is given by the scalar density operator $\bar{q}q$. The matching with the nucleon effective coupling is linked to the generation of nucleon masses \cite{Fan:2010gt}. From trace anomaly, the mass of the nucleons is given by
\begin{equation}
m_{N}=
\langle N|\left[ \sum_{q}m_{q}\bar{q}q \, + \, \frac{\beta}{4\alpha_{s}}G_{\mu\nu}G^{\mu\nu} \right]|N\rangle
\end{equation}
where the $\beta$ function at lowest order is $\beta=-\alpha_{s}^{2}/2\pi (11-2n_{f}/3)$ and $\alpha_s$ is the strong coupling constant. The heavy quark fields $Q=c,b,t$ can be integrated out trough the expansion \cite{Shifman:1978zn}
\begin{equation}
m_{Q}\bar{Q}Q\rightarrow-\frac{2}{3}\frac{\alpha_{s}}{8\pi}G_{\mu\nu}G^{\mu\nu} \ ,
\end{equation}
so that
\begin{equation}
m_{N}=\langle N|\left[ \sum_{q=u,d,s}m_{q}\bar{q}q \, - \, \frac{9\alpha_{s}}{8\pi}G_{\mu\nu}G^{\mu\nu} \right]|N\rangle \ .
\end{equation}
We can now define the fractions of nucleon mass as
\begin{equation}
\begin{split}
f_{Tq}^{(N)}=\frac{\langle N|m_{q}\bar{q}q|N\rangle}{m_{N}},\\
f_{TG}^{(N)}=1-\sum_{q=u,d,s}f_{Tq}^{(N)}.
\end{split}
\end{equation}
We consider a scalar interaction term with the $X$ particle defined by
\begin{equation}
{\cal L} = X\sum_{q}C_{S}^{q}\frac{m_{q}}{v}\bar{q}q \, + \, C_{S}^{g}\frac{\alpha_{s}}{8\pi v}XG_{\mu\nu}G^{\mu\nu} \ ,
\end{equation}
where $v=246$ GeV is the Higgs vacuum expectation value. The last one is an effective interaction term that can be generated at loop level by massive particles in the low energy limit. The nucleon effective interaction reads
\begin{equation}
{\cal L} = X\sum_{N=p,n}z_{N}\bar{N}N
\end{equation}
where
\begin{equation}
z_{N}=\frac{m_{N}}{v}\left[ \sum_{q=u,d,s}C_{S}^{q}f_{Tq}^{(N)} \, - \, \frac{1}{9}f_{TG}^{(N)}\left(C_{S}^{g}-\frac{2}{3}\sum_{q=c,b,t}C_{S}^{q}\right) \right]
\end{equation}
are the effective scalar couplings of the nucleons.
The values of the fractions of nucleon mass are given by \cite{Ellis:2000ds}
\begin{align}
& f_{Tu}^{(p)}=0.020\pm0.004,\quad f_{Td}^{(p)}=0.026\pm0.005 ,\quad f_{Ts}^{(p)}=0.118\pm0.062 \nonumber \ ,\\
& f_{Tu}^{(n)}=0.014\pm0.003,\,\,f_{Td}^{(n)}=0.036\pm0.008 ,\quad f_{Ts}^{(n)}=0.118\pm0.062 \ .
\end{align}

\subsection*{Pseudoscalar interaction}

The pseudoscalar density $i\bar{q}\gamma^{5}q$ is proportional to the divergence of the axial current $\bar{q}\gamma^{\mu}\gamma^{5}q$. The matching with the nucleon effective operator is then done with the same  $\Delta_{q}^{(N)}$ parameters already used. For the light quark contribution, we have
\begin{equation}
\langle N|m_{q}i\bar{q}\gamma^{5}q|N\rangle=m_{N}\Delta_{q}^{(N)}\,-\,\langle N|\frac{\alpha_{s}}{8\pi}G_{\mu\nu}\tilde{G}^{\mu\nu}|N\rangle \ ,
\end{equation}
while for the heavy quark fields it's enough to expand them as \cite{Shifman:1978zn}
\begin{equation}
m_{Q}i\bar{Q}\gamma^{5}Q\rightarrow-\frac{\alpha_{s}}{8\pi}G_{\mu\nu}\tilde{G}^{\mu\nu} \ .
\end{equation}
The nucleon matrix element for the pseudoscalar gluon operator is given by~\cite{Cheng:2012qr}
\begin{equation}
\langle N|\frac{\alpha_{s}}{8\pi}G_{\mu\nu}\tilde{G}^{\mu\nu}|N\rangle=m_{N}\bar{m}\left( \frac{\Delta_{u}^{(N)}}{m_{u}} + \frac{\Delta_{d}^{(N)}}{m_{d}} + \frac{\Delta_{s}^{(N)}}{m_{s}} \right) \ ,
\end{equation}
where $\bar{m}^{-1}=m_{u}^{-1}+m_{d}^{-1}+m_{s}^{-1}$.
By considering an interaction term for quarks and gluons given by
\begin{equation}
{\cal L} = X\sum_{q}C_{P}^{q}\frac{m_{q}}{v}i\bar{q}\gamma^{5}q \, - \, C_{P}^{g}\frac{\alpha_{s}}{8\pi v}XG_{\mu\nu}\tilde{G}^{\mu\nu} \ ,
\end{equation}
which is equivalent to
\begin{equation}
{\cal L}=-\frac{\partial_{\mu}X}{2v}\sum_{q}C_{P}^{q}\bar{q}\gamma^{\mu}\gamma^{5}q \, - \, C_{P}^{gg}\frac{\alpha_{s}}{8\pi v}XG_{\mu\nu}\tilde{G}^{\mu\nu} \ ,
\end{equation}
with $C_P^{gg} = C_P^g + \sum_{q} C_P^q$, the effective nucleon interaction then reads
\begin{equation}
{\cal L} = X\sum_{N=p,n}h_{N}i\bar{N}\gamma^{5}N \ ,
\end{equation}
where
\begin{equation}
\label{abs}
h_{N}=\frac{m_{N}}{v}\sum_{q=u,d,s}\Delta_{q}^{(N)}\left(C_{P}^{q}-\frac{\bar{m}}{m_{q}}C_{P}^{gg}\right)
\end{equation}
are the effective pseudoscalar couplings of the neutrons. 
%%%%%%%%%%%%%%%%%%%%%%%%%%%%%%%%%%%%%%%%%%%%%%%%%%%%%%%%%%%%%

\section{Combined analysis with both $^8$Be(18.15) and $^8$Be(17.64) energy levels}\label{appBev}

As discussed in Sec.~\ref{sec:res} in a later publication~\cite{Krasznahorkay:2017gwn} the ATOMKI collaboration reported the observation of the anomalous signal also in the $^8$Be(17.64) transition, which was absent in their first analysis~\cite{Krasznahorkay:2015iga,Krasznahorkay:2018snd}.
In this section we show how our results are modified by considering both the 
the$^8$Be(18.15) and $^8$Be(17.64) excited states. The best fit value for the anomalous decay rate for the $^8$Be(17.64) transition is~\cite{Krasznahorkay:2017gwn}
\be
\label{BeBestFit2}
\frac{\Gamma({^8 \text{Be}}(17.64) \to {^8 \text{Be}} \, + \, X)}{\Gamma({^8 \text{Be}}(17.64) \to {^8 \text{Be}} \, + \, \gamma)}\ \text{BR}(X\to e^{+} e^{-})=4.0\times10^{-6} \ .
\ee
We will associate a relative error to this best fit, not provided by ATOMKI collaboration, equal to the one from the $^8$Be(18.15) measurement of Eq.~\eqref{BeBestFit}. 

\begin{figure}[t!]
\centering
{\includegraphics[width=.55\textwidth]{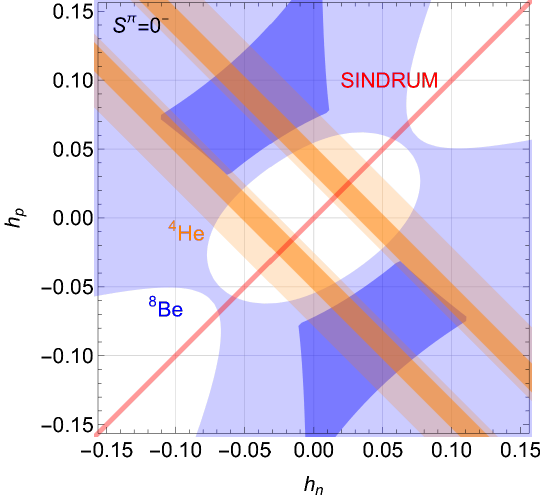}}
\caption{Regions of the $h_{n,p}$ effective nuclear couplings of a pure pseudoscalar states where the $^{8}$Be (blue) and $^{4}$He (orange) anomalous ATOMKI transition can be explained at $1\sigma$ or $2\sigma$. Inside the red region the SINDRUM bound is satisfied. Here both the $^8$Be(18.15) and the $^8$Be(17.64) transitions are considered.}
\label{plot:pseudo-v}
\end{figure}

\begin{figure}[t!]
\centering
{\includegraphics[width=.49\textwidth]{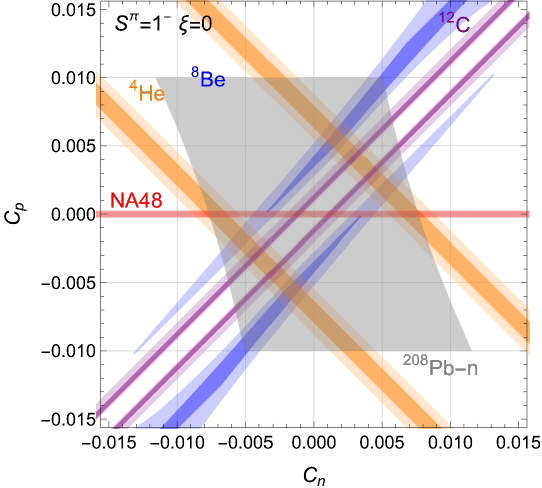}}\hfill
{\includegraphics[width=.49\textwidth]{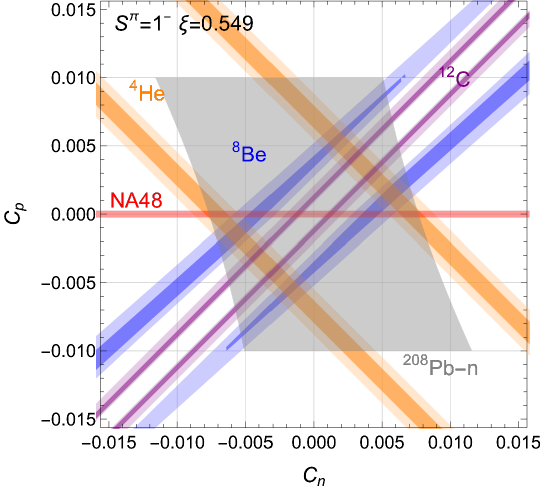}}\\
\vskip 6pt
{\includegraphics[width=.496\textwidth]{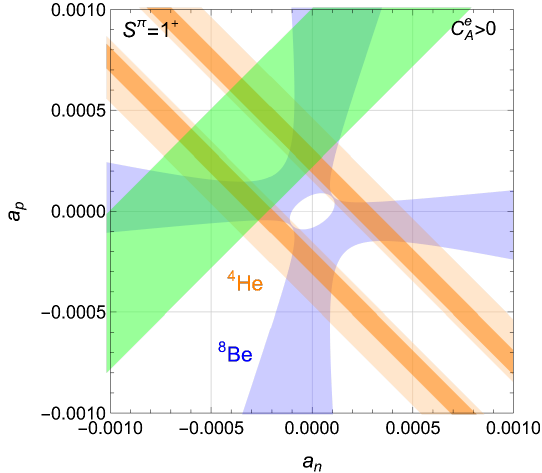}} \hfill
{\includegraphics[width=.496\textwidth]{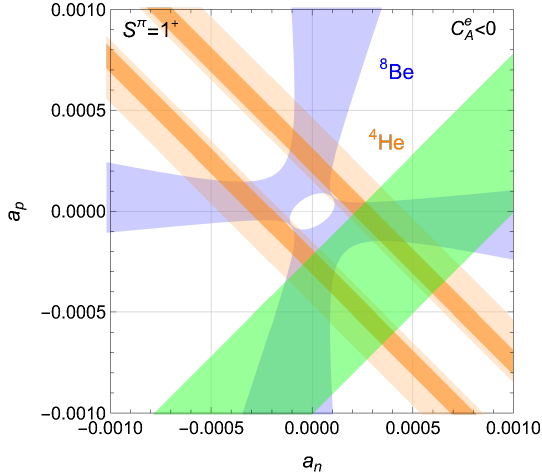}} 
\caption{{\it Upper panels:} Regions of the $C_{n,p}$ effective nuclear couplings of a pure vector state where the $^{8}$Be (blue), $^{4}$He (orange) and $^{12}$C (purple) anomalous ATOMKI transition can be explained at $1\sigma$ or $2\sigma$. Inside the red and the gray region, respectively, the NA48 and the $^{208}$Pb-n scattering bound are satisfied. In the left and right panel we assume $\xi=0$ and $\xi=0.549$ respectively, see App.~\ref{sec:em_ME} for details.
{\it Lower panels:}  Regions of the $a_{n,p}$ effective nuclear couplings of a pure axial vector state where the $^{8}$Be (blue) and $^{4}$He (orange) anomalous ATOMKI transition can be explained at $1\sigma$ or $2\sigma$. In the green region the KTeV anomaly in $\pi \to e^+ e^-$ decay can be satisfied, by assuming a positive (left panel) and negative (right panel) value for the $C_A^e$ axial coupling of the $X$ boson to electrons that can explain the anomalous $(g-2)_e$. In all figures both the $^8$Be(18.15) and the the $^8$Be(17.64) transitions are considered, see main text for more details.}
\label{plot:vector-v}
\end{figure}

\subsection*{Pseudoscalar scenario}

We summarize the results for the pure pseudoscalar scenario in Fig.~\ref{plot:pseudo-v}, where the shaded blue and orange areas represent the $1\sigma$ and $2\sigma$ compatibility regions with the ATOMKI $^8$Be and $^4$He anomalies respectively, where the former is a combination arising from both the $^8$Be energy levels.
We also overlay in red the region of parameter space satisfying the SINDRUM bound from $\pi^+ \to e^+ \nu_e X$ decay~\cite{SINDRUM:1986klz,Alves:2017avw}.
Note that a combined explanation of the $^8$Be and $^4$He anomalies is not anymore possible, once we include the constraint from the $^8$Be(17.64) transition.

\subsection*{Vector and axial scenarios}

We summarize the results for the spin-1 cases in Fig.~\ref{plot:vector-v},
with the same color code as Fig.~\ref{plot:pseudo-v} for the regions satisfying the ATOMKI anomalies. 
In the upper panels we show the results for the $S^\pi = 1^-$ assignment for the $X$ boson. As it can be seen for both $\xi$ assignments, a combined explanation of the $^8$Be, blue region, and $^4$He, orange region, anomalies at $1\sigma$ is in tension with the NA48 constraint,
while it is possible at the $2\sigma$ level. The $1\sigma$ and $2\sigma$ bands related to the $^{12}$C transition are shown in purple in the upper panels of Fig.~\ref{plot:vector}.
Note that, if confirmed, the $^{12}$C ATOMKI anomaly is in tension with a combined explanation of the $^8$Be and $^4$He anomalies and the {\emph{protophobia}} constraint.
On the other side an axial vector $S^\pi = 1^+$ state can explain both the $^8$Be and $^4$He ATOMKI anomalies at 2$\sigma$, as shown in the lower panels of Fig.~\ref{plot:vector-v}, with axial couplings to the nucleon of ${\cal O}(10^{-4})$.
Thus the inclusion of the $^8$Be(17.64) transition does not change drastically the conclusion for the spin-1 cases.

%%%%%%%%%%%%%%%%%%%%%%%%%%%%%%%%%%%%%%%%%%%%%%%%%%%%%%%%%%%%%

\section{Cross section for resonance production}\label{NWAApp}

For resonance production $p+A\rightarrow N^{*}$, the unpolarized cross section expression is given by
\begin{align}
&\sigma(p+A\rightarrow N^{*})= \nonumber \\
&=\frac{1}{(2J_{p}+1)(2J_{A}+1)}\frac{1}{4m_{A}E_{p}v_{p}}\int \frac{d^{3}p_{*}}{(2\pi)^{3}2E_{*}}(2\pi)^{4}\delta(p_{*}-p_{A}-p_{p})\sum_{pol.}|\mathcal{M}(p+A\rightarrow N^{*})|^{2}= \nonumber\\
&=\frac{1}{(2J_{p}+1)(2J_{A}+1)}\frac{(2\pi)\delta(E_{CM}^{2}-m_{*}^{2})}{4m_{A}E_{p}v_{p}}\sum_{pol.}|\mathcal{M}(p+A\rightarrow N^{*})|^{2} \ ,
\end{align}
where $E_{CM}^{2}=(p_{A}+p_{p})^{2}$.
Here we have treated the excited state $N^{*}$ as a bound states of $p$ and $A$ with mass $m_{*}$ and spin $J_{*}$. The reverse process defines the decay width
\begin{align}
&\Gamma(N^{*}\rightarrow p+A)= \nonumber \\
&=\frac{1}{(2J_{*}+1)}\frac{1}{2m_{*}}\int \frac{d^{3}p_{p}}{(2\pi)^{3}2E_{p}}\int \frac{d^{3}p_{A}}{(2\pi)^{3}2E_{A}}(2\pi)^{4}\delta(p_{*}-p_{A}-p_{p})\sum_{pol.}|\mathcal{M}(N^{*}\rightarrow p+A)|^{2}= \nonumber \\
&=\frac{1}{(2J_{*}+1)}\frac{\sqrt{\lambda(m_{*}^{2},m_{A}^{2},m_{p}^{2})}}{16\pi m_{*}^{3}}\sum_{pol.}|\mathcal{M}(N^{*}\rightarrow p+A)|^{2}
\end{align}
where $\lambda(x,y,z)=x^{2}+y^{2}+z^{2}-2xy-2xz-2yz$. The angular integration is trivial because the unpolarized squared matrix element of three momentum is just a function of the masses.
Since the nuclear force, which mediate the interaction, is invariant under time reversal, it follows that 
\begin{equation}
\mathcal{M}(N^{*}\rightarrow p+A)=\mathcal{M}(p+A\rightarrow N^{*}) \ .
\end{equation}
By that, one has
\begin{equation}
\sigma(p+A\rightarrow N^{*})=\frac{(2J_{*}+1)}{(2J_{p}+1)(2J_{A}+1)}\frac{8\pi^{2} m_{*}^{3}}{m_{A}E_{p}v_{p}}\frac{\Gamma(N^{*}\rightarrow p+A)}{\sqrt{\lambda(m_{*}^{2},m_{A}^{2},m_{p}^{2})}}\delta(E_{CM}^{2}-m_{*}^{2}) \ .
\end{equation}
For an off-shell state, since the excited nucleus is unstable, one should broaden the $\delta$-function into a resonance peak by taking the narrow width approximation
\begin{equation}
\delta(E_{CM}^{2}-m_{*}^{2})\rightarrow\frac{1}{\pi}\frac{m_{*}\Gamma_{*}}{(E_{CM}^{2}-m_{*}^{2})^{2}+m_{*}^{2}\Gamma_{*}^{2}} \ ,
\end{equation}
finally obtaining
\be
\label{resprod}
\sigma(p+A\rightarrow N^{*})=\frac{(2J_{*}+1)}{(2J_{p}+1)(2J_{A}+1)}\frac{8\pi m_{*}^{3}}{m_{A}E_{p}v_{p}}\frac{\Gamma(N^{*}\rightarrow p+A)}{\sqrt{\lambda(m_{*}^{2},m_{A}^{2},m_{p}^{2})}}\frac{m_{*}\Gamma_{*}}{(E_{CM}^{2}-m_{*}^{2})^{2}+m_{*}^{2}\Gamma_{*}^{2}}.
\ee

%%%%%%%%%%%%%
%%%%  EQS  %%%%%
%%%%%%%%%%%%%

\section{Experimental constraints on a spin-1 boson}
\label{sec_pheno}

A light boson coupled to first generation quarks and leptons is subject to a large variety of experimental constraints. In this section we recap the most relevant ones that affects a possible explanation of the ATOMKI anomaly through a BSM degree of freedom with mass $\sim 17\;$MeV for the spin-1 case.
We parametrize the effective UV interactions of a spin-1 state $X_\mu$  with leptons and quarks as
\be\label{eq:lagV}
\begin{split}
{\cal L} = &X_\mu \sum_{f=q,l...} \bar\psi^f (C^f_V + \gamma^5 C^f_A ) \psi^f 
= X_\mu \sum_{f=q,l...} \left( C^f_{L,R} \bar\psi^f_{L,R} \gamma^\mu \psi^f_{L,R} \right)  \ ,
\end{split}
\ee
where $C^f_{V,A} =\frac{1}{2}(C^f_R\pm C^f_L)$
and we assume  diagonal couplings in flavor space. The connection between the quark and nucleon couplings can be found in App. \ref{qton}. We list in the following the most relevant constraints for the spin-1 case coupling to the first generation of quarks and leptons.

\subsubsection*{$e^+ e^- \to \gamma X$ scattering}

By neglecting kinematic differences with respect to the pure dark photon case, we can recast the bound from the
KLOE experiment at the DA$\Phi$NE collider~\cite{Anastasi:2015qla} from which we obtain a bound
\be
\sqrt{(C_V^e)^2+(C_A^e)^2} \lesssim \frac{6.1 \times 10^{-4}}{\sqrt{{\rm BR}(X\to e^+ e^-)}} \ .
\ee

\subsubsection*{Parity violation}

Parity violation in M\o ller scattering constraints the product of the vector and axial couplings.
The most sensitive measurement arises from SLAC E158~\cite{Anthony:2005pm} at $Q^2 = 160\;$MeV$^2$. The measurement in~\cite{Anthony:2005pm} has been recast in~\cite{Kahn:2016vjr} and the obtained bound reads
\be
|C_V^e \times C_A^e| \lesssim 10^{-8} \ .
\ee

\subsubsection*{Beam dump experiments}

Beam dump experiments look for $X$ production via bremsstrahlung from electrons scattering off target nuclei. For the $X$ particle 
not to be seen in these experiments there are two possibilities: either the particle is not produced at all, or its decay products are caught in the dump, thus setting both an upper and lower limit for the couplings of the $X$ boson. In the first case the stronger limit comes from the E137 experiment~\cite{Bjorken:1988as}, see also~\cite{Andreas:2012mt},  which is independent on the $X$ decay rate and gives
\be
\sqrt{(C_V^e)^2+(C_A^e)^2} \lesssim 1.1 \times 10^{-8} \ ,
\ee
while in the second case the stronger limit comes from the NA64 experiment~\cite{Banerjee:2018vgk,Banerjee:2019hmi} for which we have
\be
\sqrt{(C_V^e)^2+(C_A^e)^2}\gtrsim 3.6 \times10^{-5} \times \sqrt{{\rm BR}(X\to e^+ e^-)}  \ .
\ee

\subsubsection*{Prompt decay in ATOMKI detector}

The requirement of a prompt decay into the ATOMKI detector imposes now the constraint
\be
\sqrt{(C_V^e)^2+(C_A^e)^2}\gtrsim 3 \times10^{-7} \times \sqrt{{\rm BR}(X\to e^+ e^-)} \ ,
\ee
which is weaker than the bound from NA64.

\subsubsection*{Atomic parity violation}

In atomic system, parity violation can be observed in the case, {\emph{e.g}}, of and electric dipole transition between two atomic states with the same parity. The $X_\mu$ gives additional contributions to these transitions due to the interaction between atomic electrons and the nucleus. In the effective operator 
\be
{\cal L} \supset - \frac{1}{m_X^2} \left[ C_V^u C_A^e (\bar u \gamma^\mu u)(\bar e \gamma_\mu \gamma^5 e) + C_A^u C_V^e (\bar u \gamma^\mu \gamma^5 u)(\bar e \gamma_\mu e)  + u \leftrightarrow d\right]  \ ,
\ee
where only the $V\times A$ part have been kept,  only the $A_e \times V_{u,d}$ interaction give a relevant effect for parity violation observables. This is due to the fact this part of the interaction between the electron and the nucleus is coherent, and thus proportional to the total weak charge of the nucleus itself, while the $A_q \times V_e$ interaction adds incoherently. This effect is thus suppressed for heavy enough nuclei~\cite{Ginges:2003qt}. The BSM contribution to $A_e \times V_{u,d}$ can be expressed as a modification to the weak nuclear charge $Q_W$~\cite{Bouchiat:2004sp}
\be
\delta Q_W = -\frac{2 \sqrt 2}{ G_F} 3(Z+N) \frac{C_A^e C_V^{q,{\rm eff}}}{m_X^2} \ ,\qquad  C_V^{q,{\rm eff}} = \frac{C_V^u (2 Z +N ) + C_V^d (Z+2N)}{3(Z+N)} \ .
\ee
The most accurate prediction comes from transition of $^{133}_{55}{\rm Cs}$~\cite{Porsev:2009pr} which, combined with the SM theoretical prediction~\cite{Dzuba:2012kx}, yields~\cite{Arcadi:2019uif} $|\delta Q_W| \lesssim 0.6$
hence the bound reads
\be
\label{apv}
|C_A^e| \left| \frac{188}{399}  C_V^u  + \frac{211}{399} C_V^d \right| \lesssim 1.8\times10^{-12} \ .
\ee

\bibliographystyle{JHEP}
{\footnotesize
\bibliography{biblio}}
\end{document}